\documentclass{aa}  
\usepackage{graphicx}
\usepackage{natbib}
\bibpunct{(}{)}{,}{a}{}{;}
\usepackage{amssymb}
\usepackage{txfonts}
\begin{document}

   \title{APEX-CHAMP$^+$ high-J CO observations of low-mass young stellar objects: II. Distribution and origin of warm molecular gas}

  \author{T.A. van Kempen
          \inst{1,2}
	  \and
	  E.F. van Dishoeck\inst{1,3}
	       \and
               R. G{\"u}sten\inst{4}
               \and
               L.E. Kristensen\inst{1}
               \and
           P. Schilke\inst{4}
       \and
        M.R. Hogerheijde\inst{1}
      \and 
      W. Boland\inst{1,5}
      \and
       K.M. Menten\inst{4}
    \and
   F. Wyrowski\inst{4}}
   
          \offprints{T. A. van Kempen}

   \institute{$^1$ Leiden Observatory, Leiden University, P.O. Box 9513,
              2300 RA Leiden, The Netherlands\\	      
             $^2$ Center for Astrophysics, 60 Garden Street, Cambridge, MA 02138, USA      \\
 $^3$ Max-Planck Institut f\"ur Extraterrestrische 
             Physik (MPE), Giessenbachstr.\ 1, 85748 Garching, Germany \\
              $^4$ Max Planck Institut f\"ur Radioastronomie, Auf dem H\"ugel 69, D-53121, Bonn, Germany\\ 
             $^5$ Nederlandse Onderzoeksschool Voor Astronomie (NOVA), P.O. Box 9513, 2300 RA Leiden, The Netherlands \\
             $^6$ SRON Netherlands Institute for Space Research , P.O. Box 800, 9700 AV Groningen, The Netherlands\\
\email{tvankempen@cfa.harvard.edu}
                     }
   \date{Draft: 0.7 July 2008}
\titlerunning{The warm gas within young low-mass protostars}


\def\placeFigurePaperSevenOne{
\begin{figure*}[!ht]
\begin{center}
\includegraphics[width=450pt]{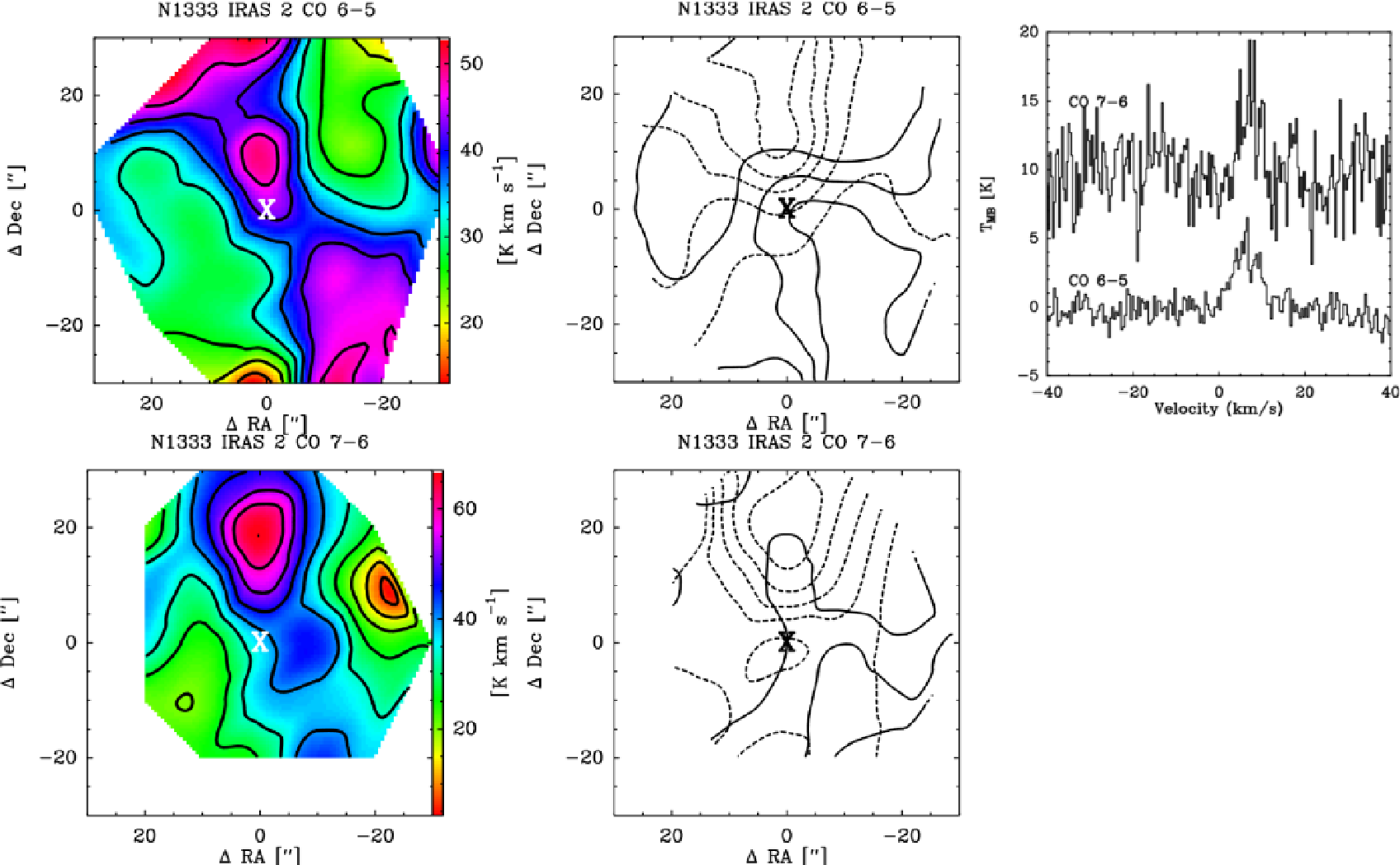}
\end{center}
\caption{
\scriptsize Maps of CO 6--5 ({\it top row}) and CO 7--6 ({\it bottom row}) of NGC 1333 IRAS 2. The left-most image shows the total integrated intensity over the line. The middle figures show the outflow contributions from the red ({\it dashed lines}) and blue ({\it solid lines}) outflow. Contours are in increasing levels of 4 K km s$^{-1}$ for both transitions. The outflow contributions are calculated by only including emission greater or smaller than +/- 1.5 km s$^{-1}$ from the central velocity. The right-most image at the top row shows the $^{12}$CO spectra at the central position. }
\label{7:fig:N1333}
\end{figure*}
}
\def\placeFigurePaperSevenTwo{
\begin{figure*}[!ht]
\begin{center}
\includegraphics[width=450pt]{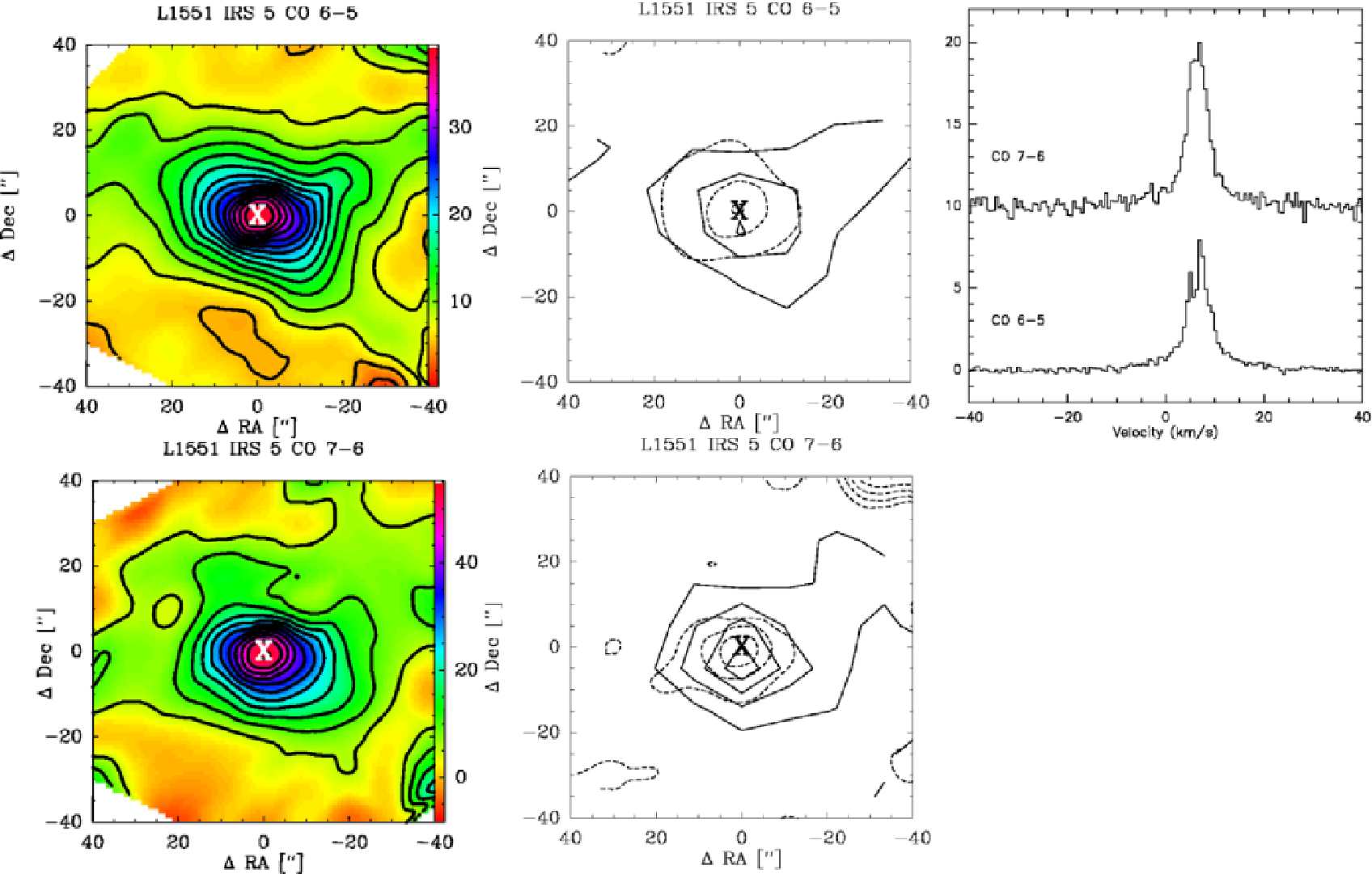}
\end{center}
\caption{\scriptsize Maps of CO 6--5 ({\it top row}) and CO 7--6 ({\it bottom row}) of L~1551 IRS 5. The left-most image shows the total integrated intensity over the line. The middle figures show the outflow contributions from the red ({\it dashed lines}) and blue ({\it solid lines}) outflow.  Contours are in increasing levels of 0.5 K km s$^{-1}$ for both transitions. The outflow contributions are calculated by only including emission greater or smaller than +/- 1.5 km s$^{-1}$ from the central velocity. The right-most image at the top row shows the $^{12}$CO spectra at the central position. }
\label{7:fig:L1551}
\end{figure*}
}

\def\placeFigurePaperSevenThree{
\begin{figure*}[!ht]
\begin{center}
\includegraphics[width=450pt]{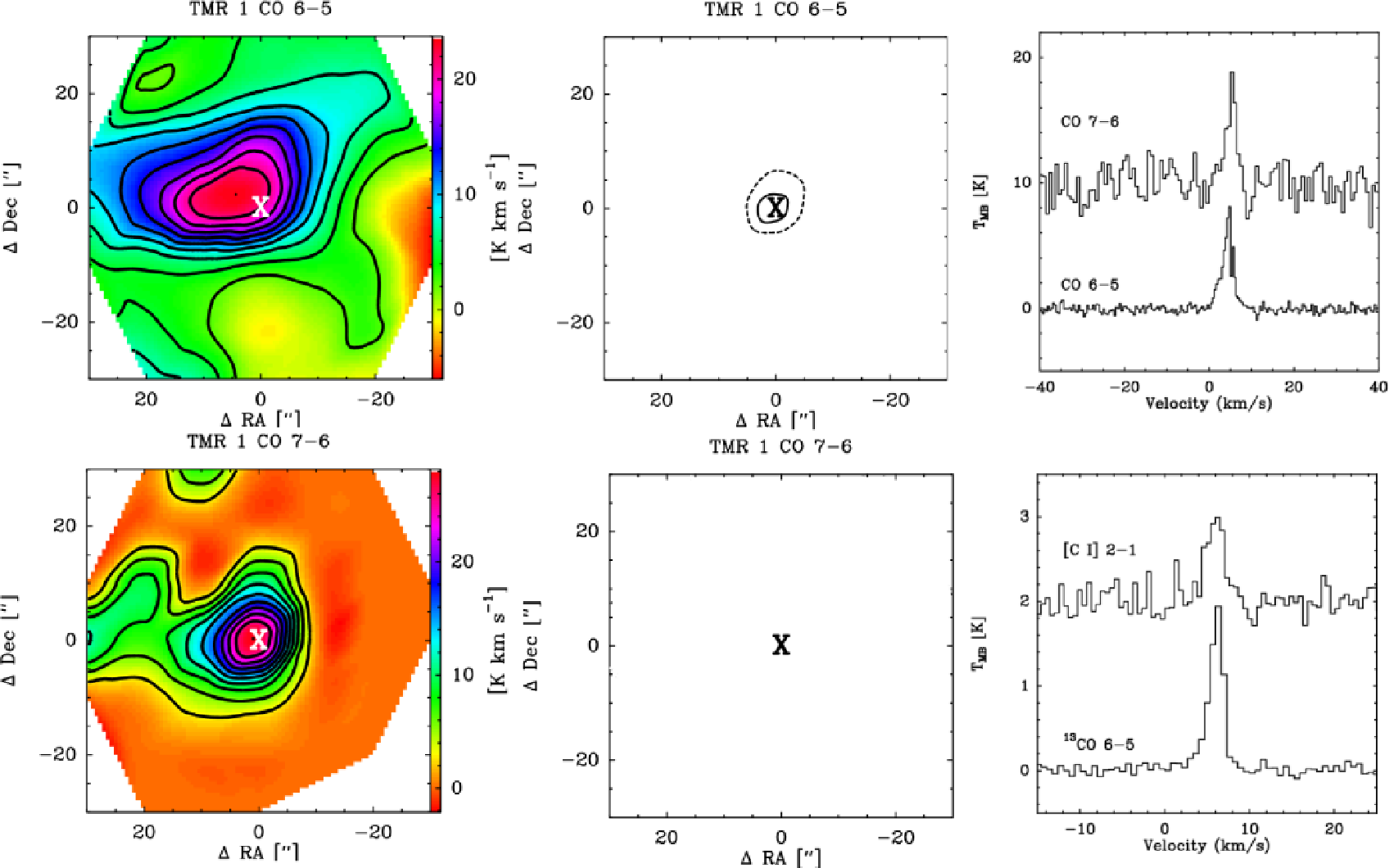}
\end{center}
\caption{\scriptsize Maps of CO 6--5 ({\it top row}) and CO 7--6 ({\it bottom row}) of TMR 1. The left-most image shows the total integrated intensity over the line. The middle figures show the outflow contributions from the red ({\it dashed lines}) and blue ({\it solid lines}) outflow. Contours are in increasing levels of 1 K km s$^{-1}$ for both transitions.The outflow contributions are calculated by only including emission greater or smaller than +/- 1.5 km s$^{-1}$ from the central velocity. The right-most image at the top row shows the $^{12}$CO spectra at the central position, while the right-most image at the bottom row shows the spectra of the observed isotopologues and [C I] 2--1 at the central position. }
\label{7:fig:TMR1}
\end{figure*}
}

\def\placeFigurePaperSevenFour{
\begin{figure*}[!ht]
\begin{center}
\includegraphics[angle=270,width=450pt]{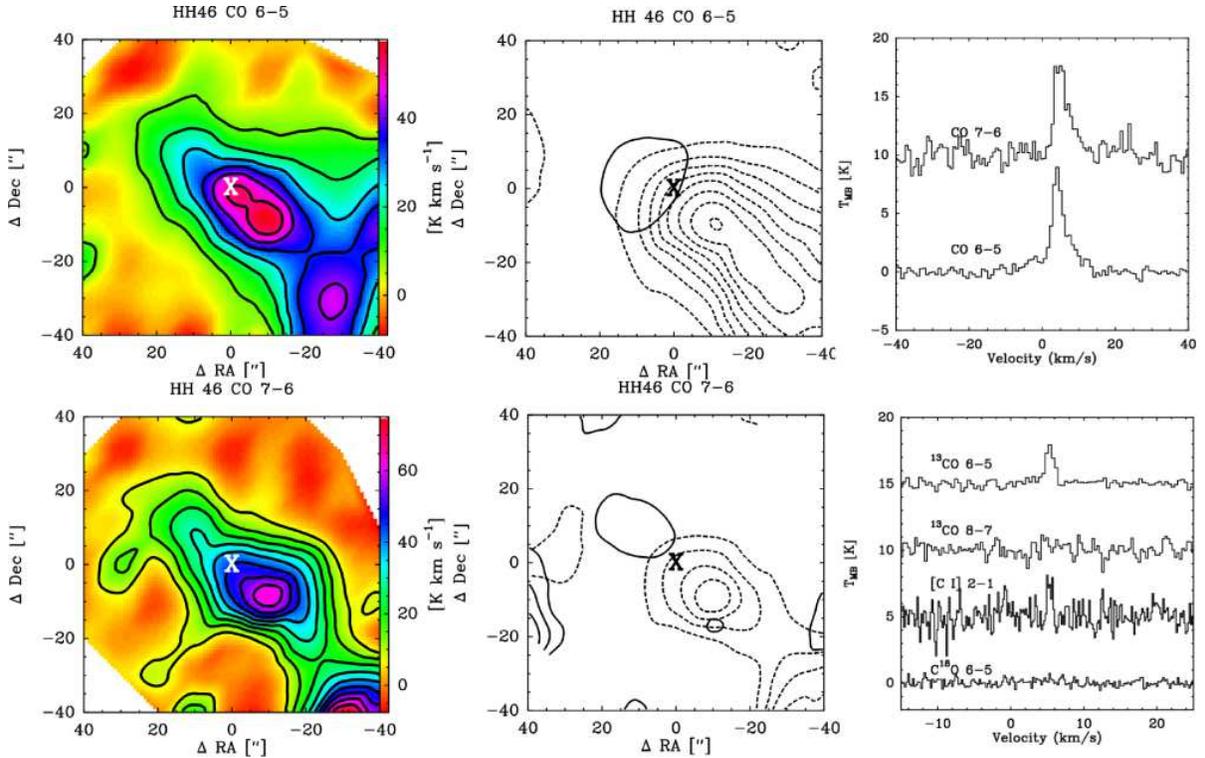}
\end{center}
\caption{\scriptsize CO 6--5 ({\it left}) CO 7--6({\it middle}) maps
  and spectra of the central position of HH 46. The left-most image
  shows the total integrated intensity over the line. The middle
  figures show the outflow contributions from the red ({\it dashed
    lines}) and blue ({\it solid lines}) outflow. Contours are in
  increasing levels of 2 K km s$^{-1}$ for CO 6--5 and 3 K km s$^{-1}$
  for CO 7--6. The outflow contributions are calculated by only
  including emission greater or smaller than +/- 1.5 km s$^{-1}$ from
  the central velocity.  The right-most image at the top row shows the
  $^{12}$CO spectra at the central position, while the right-most
  image at the bottom row shows the spectra of the observed
  isotopologues and [C I] 2--1 at the central position (see also paper I).  }
\label{7:fig:HH46}
\end{figure*}
}

\def\placeFigurePaperSevenFive{

\begin{figure*}[!ht]
\begin{center}
\includegraphics[angle=270,width=450pt]{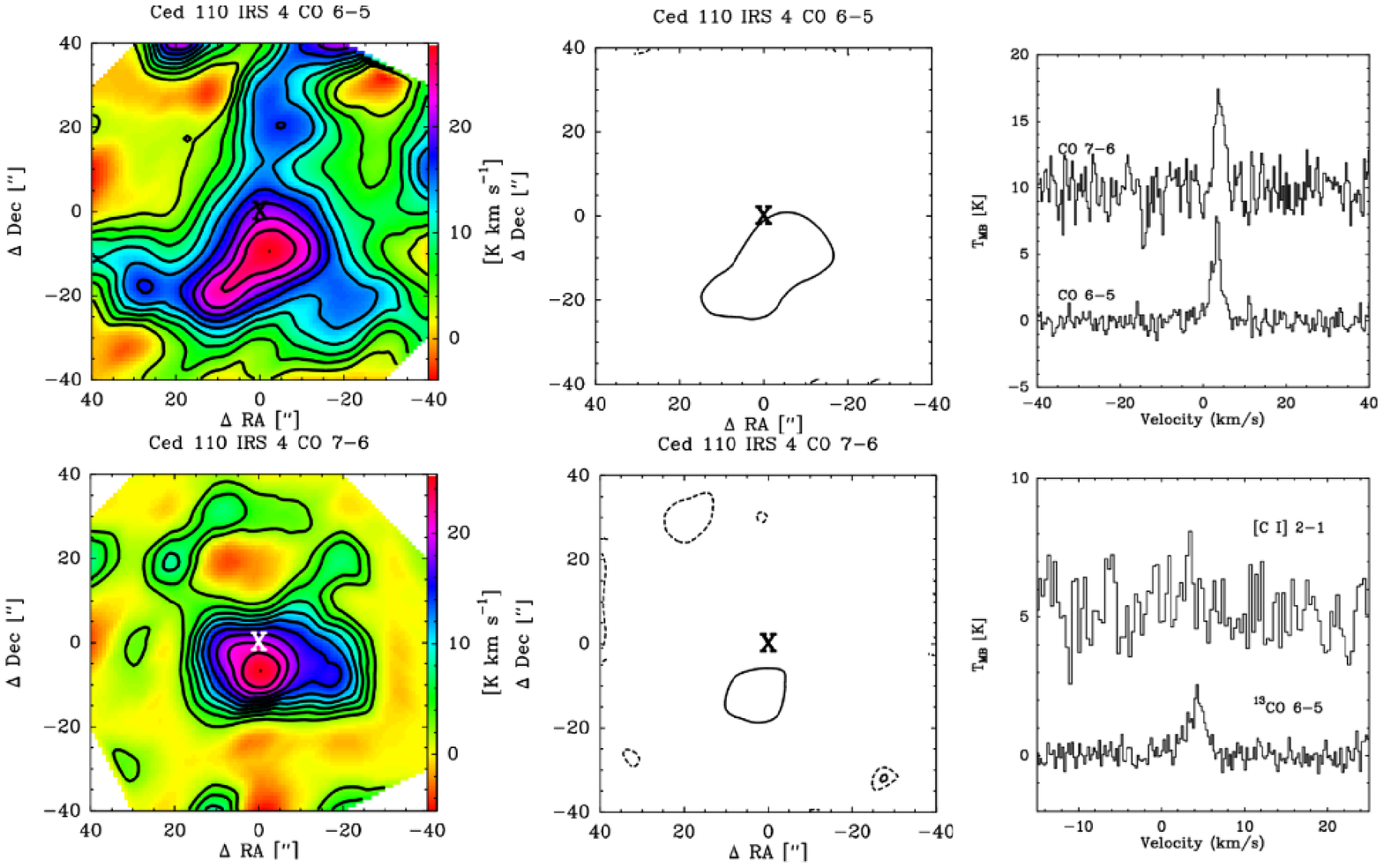}
\end{center}
\caption{\scriptsize Maps of CO 6--5 ({\it top row}) and CO 7--6 ({\it bottom row}) of Ced 110 IRS 4. The left-most image shows the total integrated intensity over the line. The middle figures show the outflow contributions from the red ({\it dashed lines}) and blue ({\it solid lines}) outflow. Contours are in increasing levels of 3 K km s$^{-1}$ for both transitions. The outflow contributions are calculated by only including emission greater or smaller than +/- 1.5 km s$^{-1}$ from the central velocity. The right-most image at the top row shows the $^{12}$CO spectra at the central position, while the right-most image at the bottom row shows the spectra of the observed isotopologues and [C I] 2--1 at the central position. }
\label{7:fig:CED110}
\end{figure*}
}

\def\placeFigurePaperSevenSix{
\begin{figure*}[!ht]
\begin{center}
\includegraphics[angle=270,width=450pt]{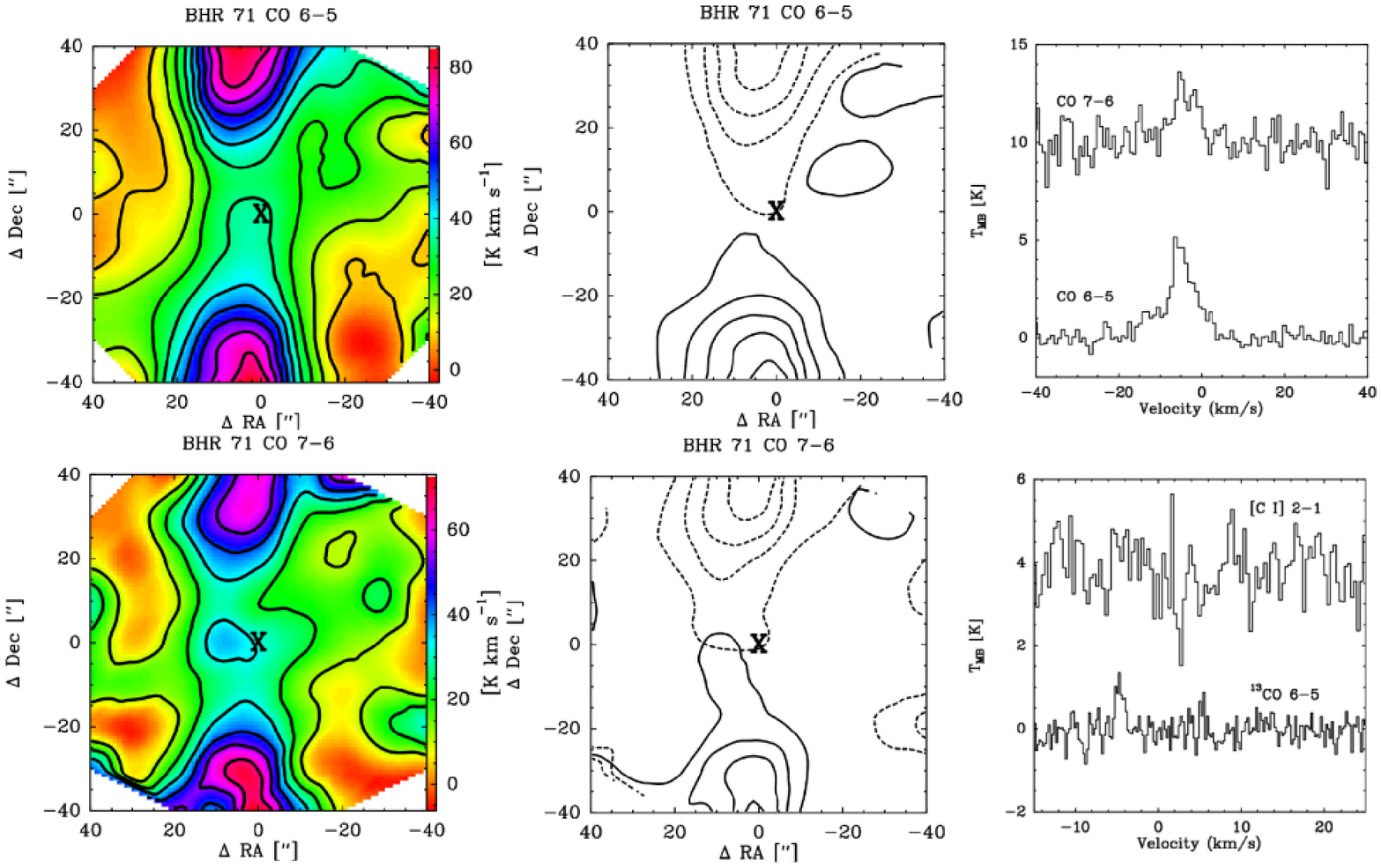}

\end{center}
\caption{\scriptsize Maps of CO 6--5 ({\it top row}) and CO 7--6 ({\it bottom row}) of BHR 71. The left-most image shows the total integrated intensity over the line. The middle figures show the outflow contributions from the red ({\it dashed lines}) and blue ({\it solid lines}) outflow. Contours are in increasing levels of 10 K km s$^{-1}$ for CO 6--5 and 5 K km s$^{-1}$ for CO 7--6. The outflow contributions are calculated by only including emission greater or smaller than +/- 1.5 km s$^{-1}$ from the central velocity. The right-most image at the top row shows the $^{12}$CO spectra at the central position, while the right-most image at the bottom row shows the spectra of the observed isotopologues and [C I] 2--1 at the central position. }
\label{7:fig:BHR71}
\end{figure*}
}

\def\placeFigurePaperSevenSeven{
\begin{figure*}[!ht]
\begin{center}
\includegraphics[angle=270,width=475pt]{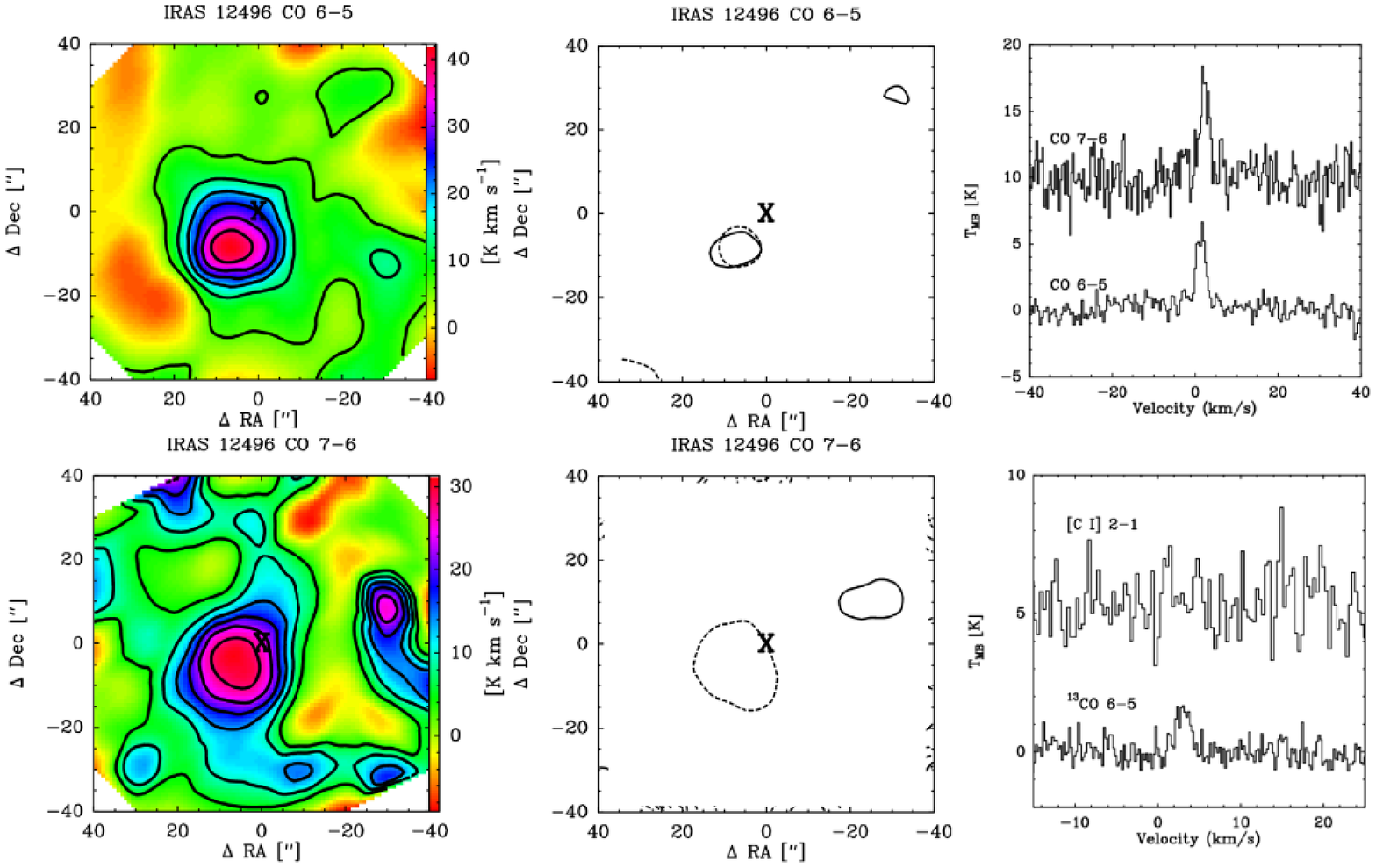}
\end{center}
\caption{\scriptsize Maps of CO 6--5 ({\it top row}) and CO 7--6 ({\it bottom row}) of IRAS 12496-7650. The left-most image shows the total integrated intensity over the line. The middle figures show the outflow contributions from the red ({\it dashed lines}) and blue ({\it solid lines}) outflow. Contours are in increasing levels of 3 K km s$^{-1}$ for both transitions. The outflow contributions are calculated by only including emission greater or smaller than +/- 1.5 km s$^{-1}$ from the central velocity. The right-most image at the top row shows the $^{12}$CO spectra at the central position, while the right-most image at the bottom row shows the spectra of the observed isotopologues and [C I] 2--1 at the central position. }
\label{7:fig:I12496}
\end{figure*}
}

\def\placeFigurePaperSevenEight{
\begin{figure*}[!ht]
\begin{center}
\includegraphics[angle=270,width=475pt]{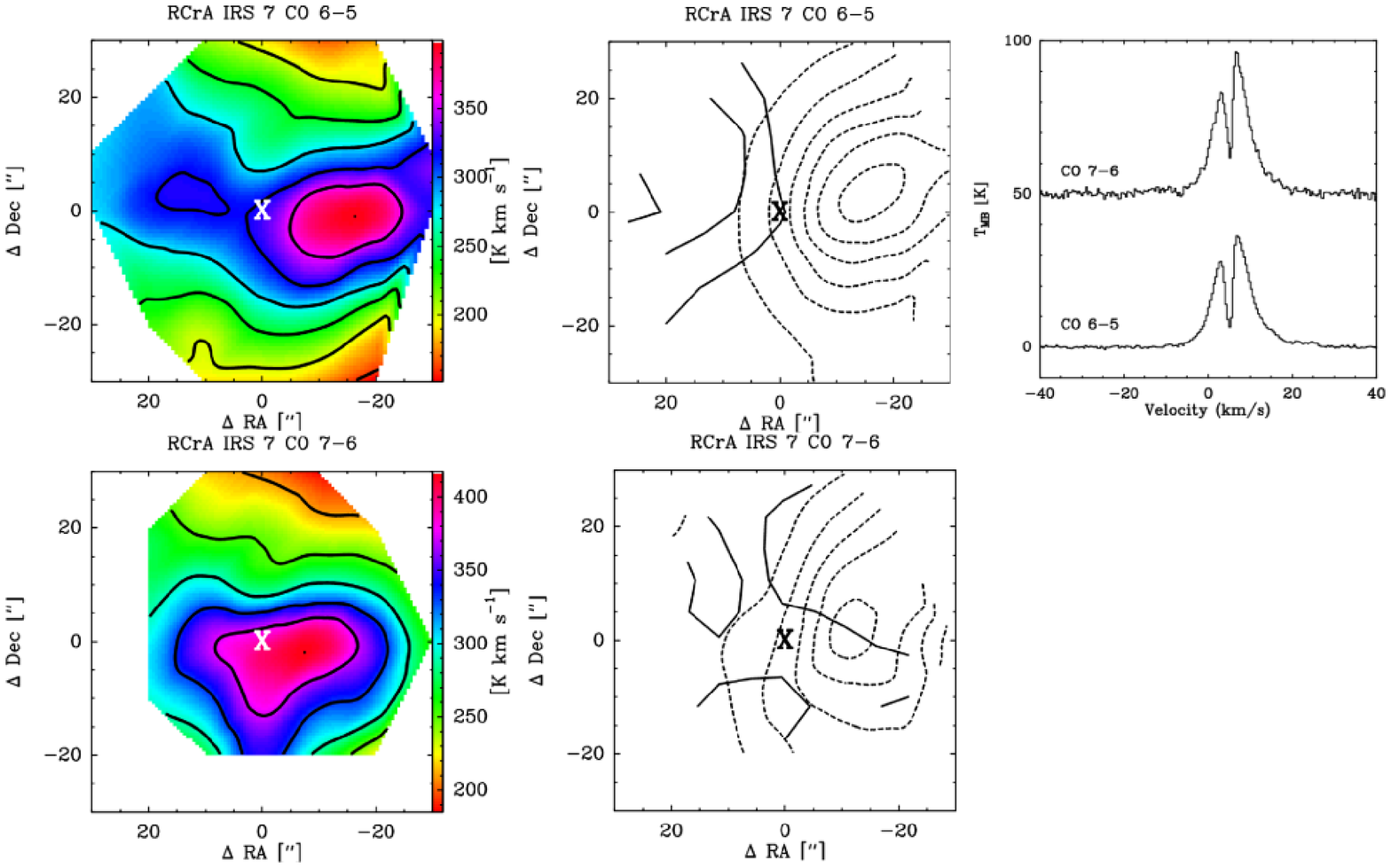}
\end{center}
\caption{\scriptsize Maps of CO 6--5 ({\it top row}) and CO 7--6 ({\it bottom row}) of RCrA IRS 7a. The left-most image shows the total integrated intensity over the line. The middle figures show the outflow contributions from the red ({\it dashed lines}) and blue ({\it solid lines}) outflow. Contours are in increasing levels of 10 K km s$^{-1}$ for both transitions. The outflow contributions are calculated by only including emission greater or smaller than +/- 8 km s$^{-1}$ from the central velocity. The right-most image at the top row shows the $^{12}$CO spectra at the central position.  }
\label{7:fig:rcra}
\end{figure*}
}

\def\placeFigurePaperSevenNine{
\begin{figure*}
\begin{center}
\includegraphics[angle=270,width=200pt]{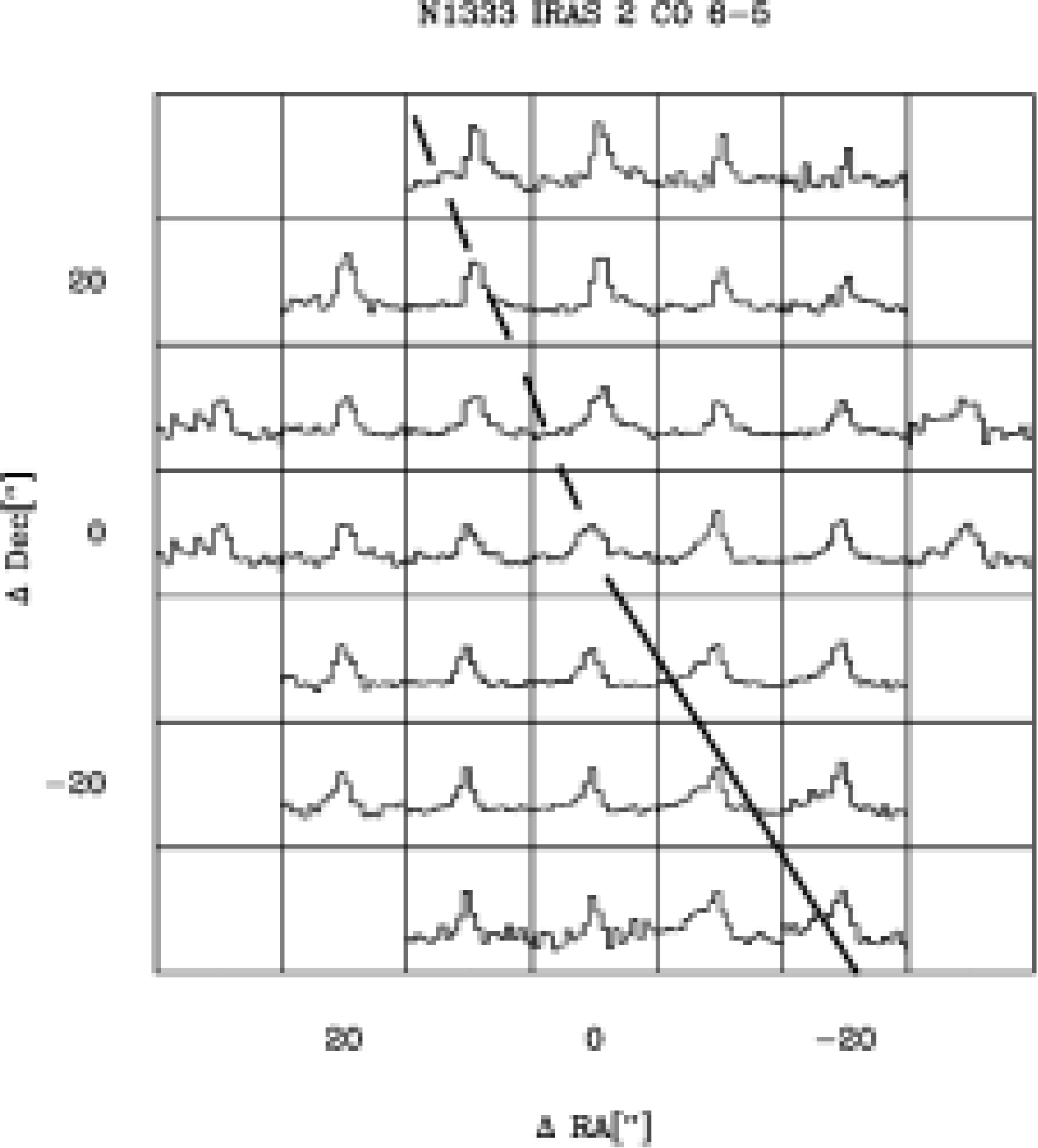}
\includegraphics[angle=270,width=230pt]{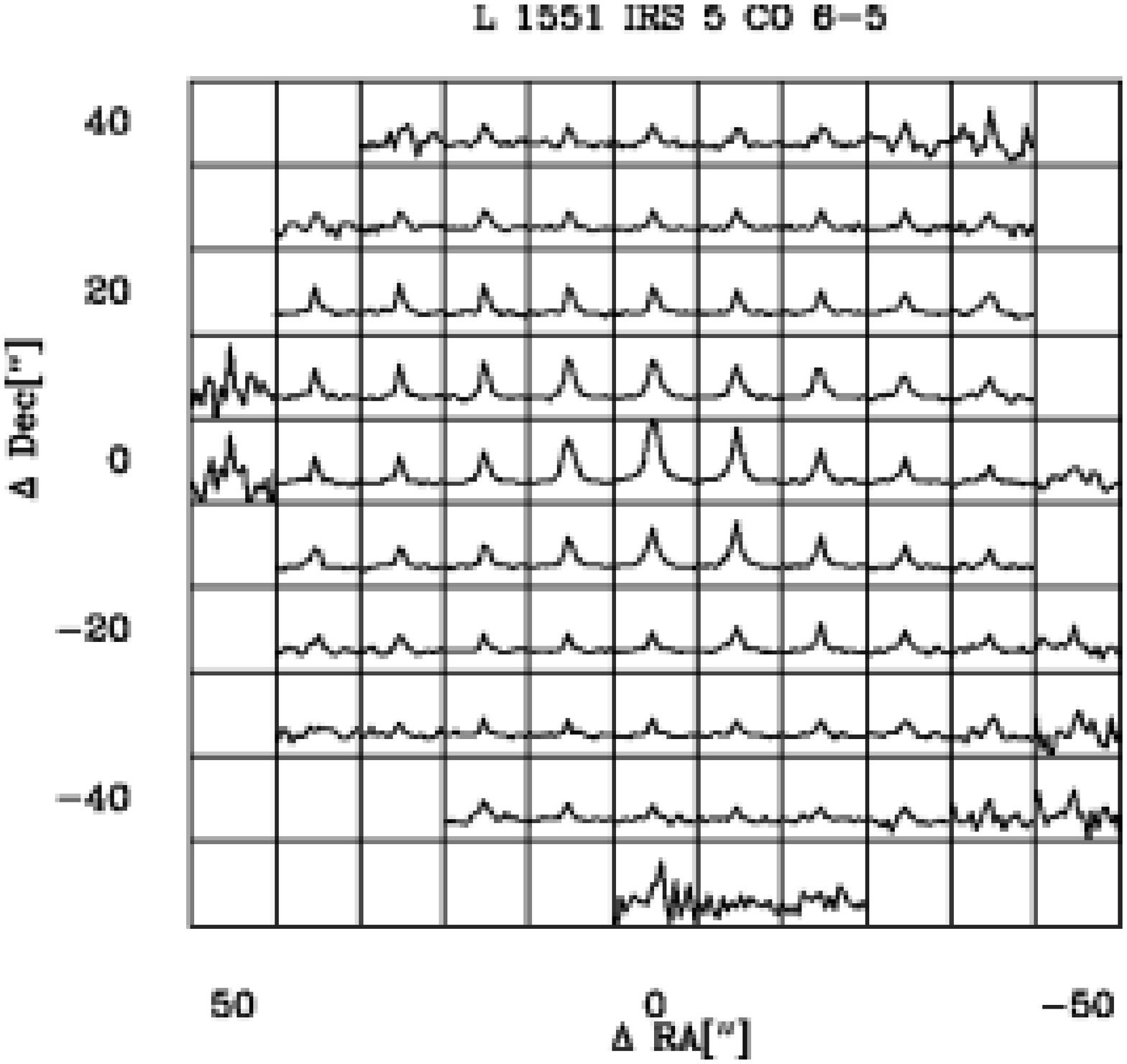}
\end{center}
\caption{Spectra of $^{12}$CO 6--5  over an area of $40''\times40''$ of NGC 1333 IRAS 2 ({\it left}) and L~1551 IRS 5 (({\it right}). NGC 1333 IRAS 2 was observed using the hexa mode, so the covered area is slightly smaller than that of L~1551 IRS 5 taken in OTF mode. For all spectra in both sources, horizontal axes range from -10 to 25 km s$^{-1}$ and vertical axes from -5 to 12 K. For both sources, the main outflow axis of the red and blue are shown with a {\it dashed} and {\it solid} lines. }
\label{7:fig:spec1}
\end{figure*}
}

\def\placeFigurePaperSevenTen{
\begin{figure*}
\begin{center}

\includegraphics[angle=270,width=200pt]{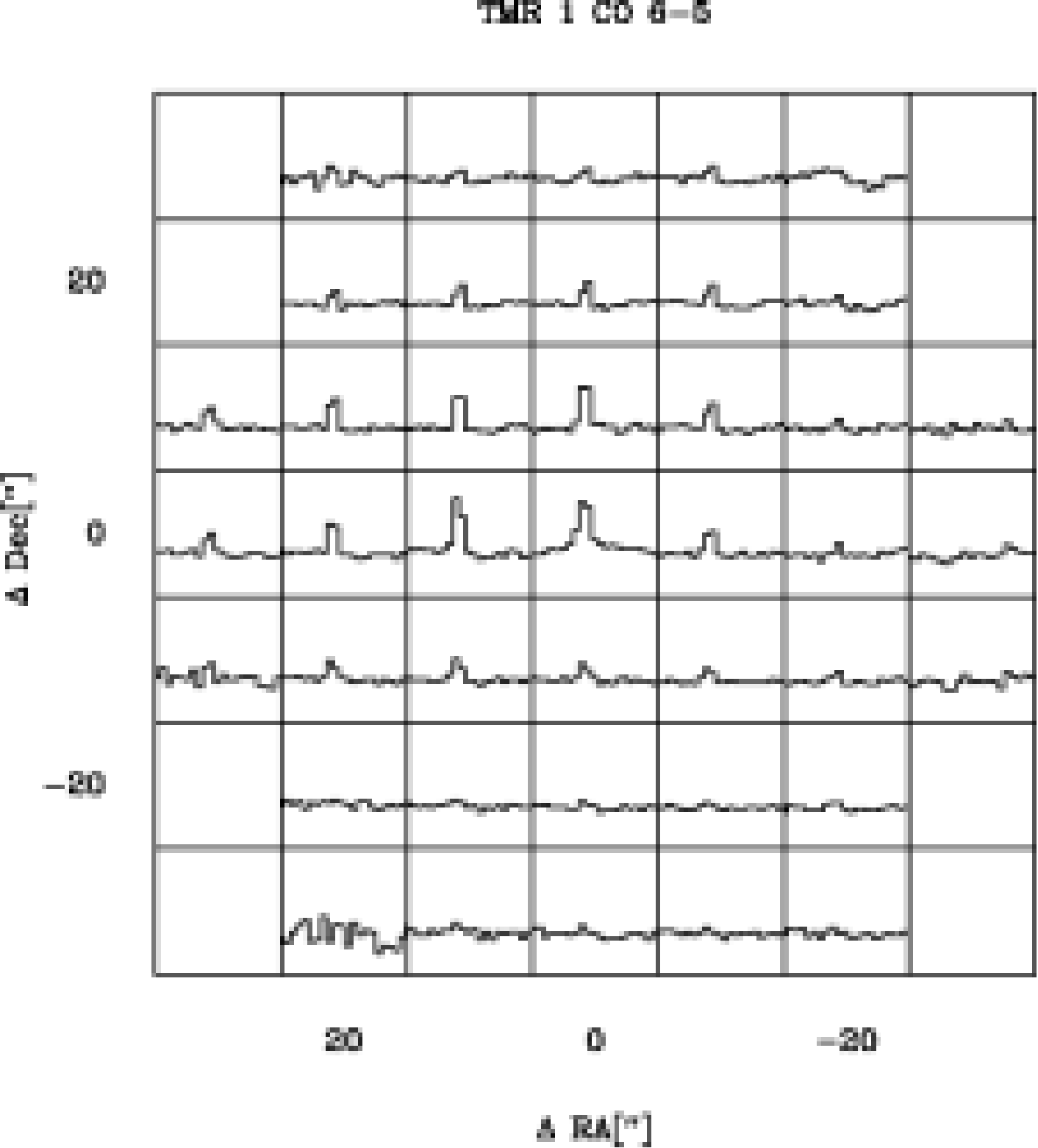}
\includegraphics[angle=270,width=200pt]{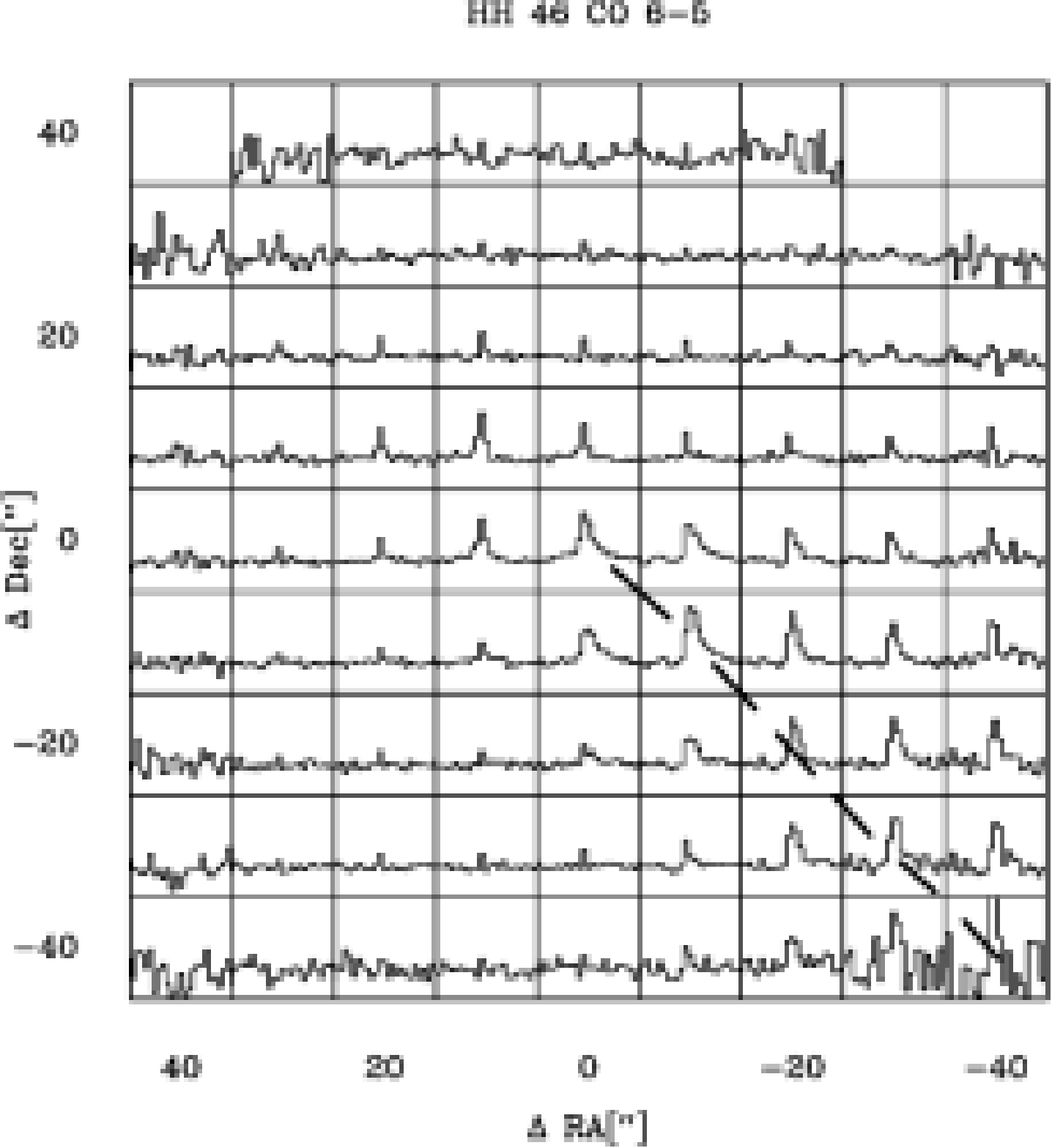}
\end{center}
\caption{Spectra of $^{12}$CO 6--5  over an area of $40''\times40''$ of TMR 1 ({\it left}) and HH 46 (({\it right}). TMR~1 was observed using the hexa mode, so the covered area is slightly smaller than that of HH~46 taken in OTF mode. For all spectra in both sources, the horizontal  axes range from -10 to 20 km s$^{-1}$ with vertical axes from -5 to 12 K. For HH~46, the red outflow axis is shown with a {\it dashed} line.}
\label{7:fig:spec2}
\end{figure*}
}

\def\placeFigurePaperSevenEleven{
\begin{figure*}
\begin{center}
\includegraphics[angle=270,width=200pt]{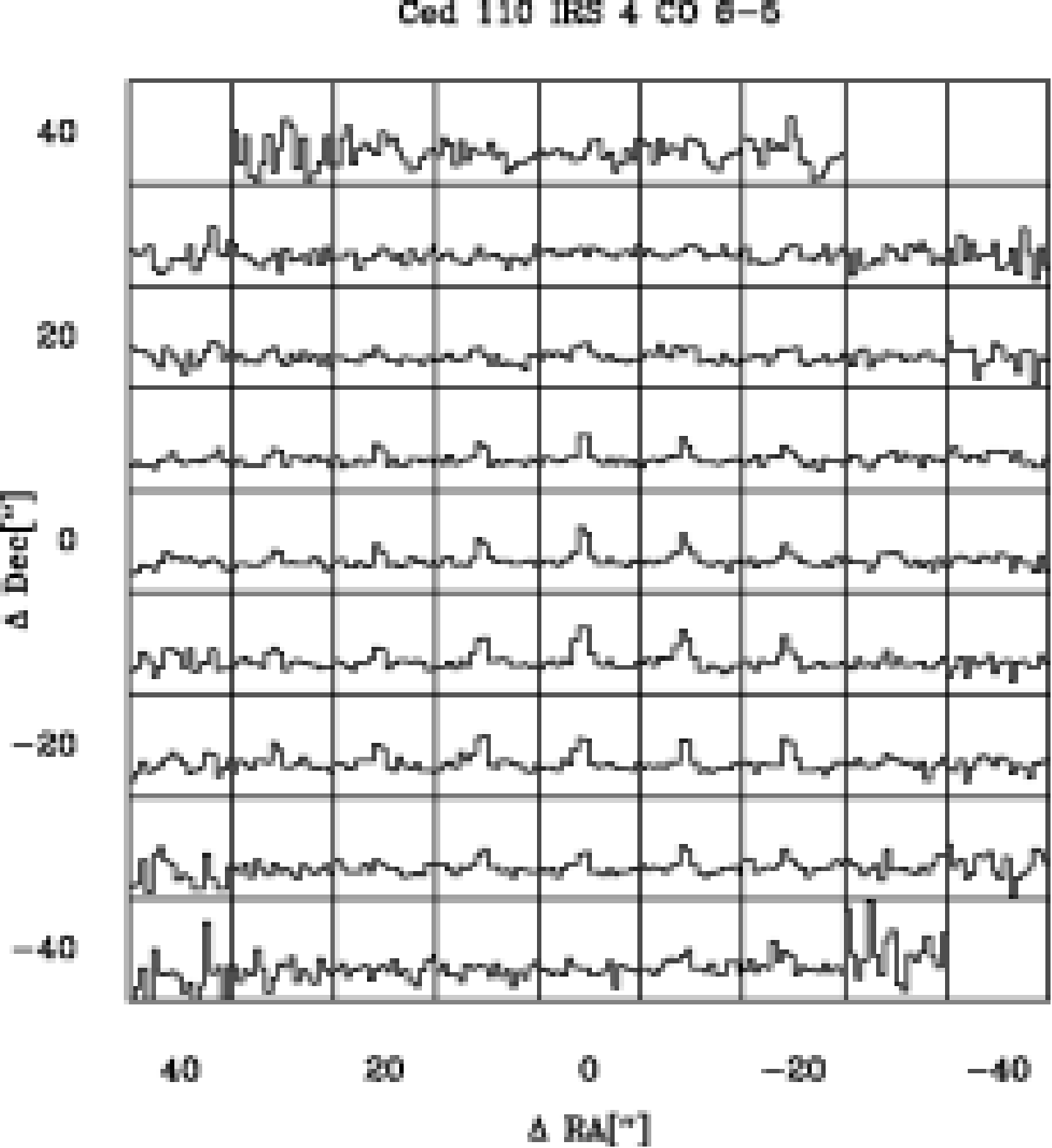}
\includegraphics[angle=270,width=200pt]{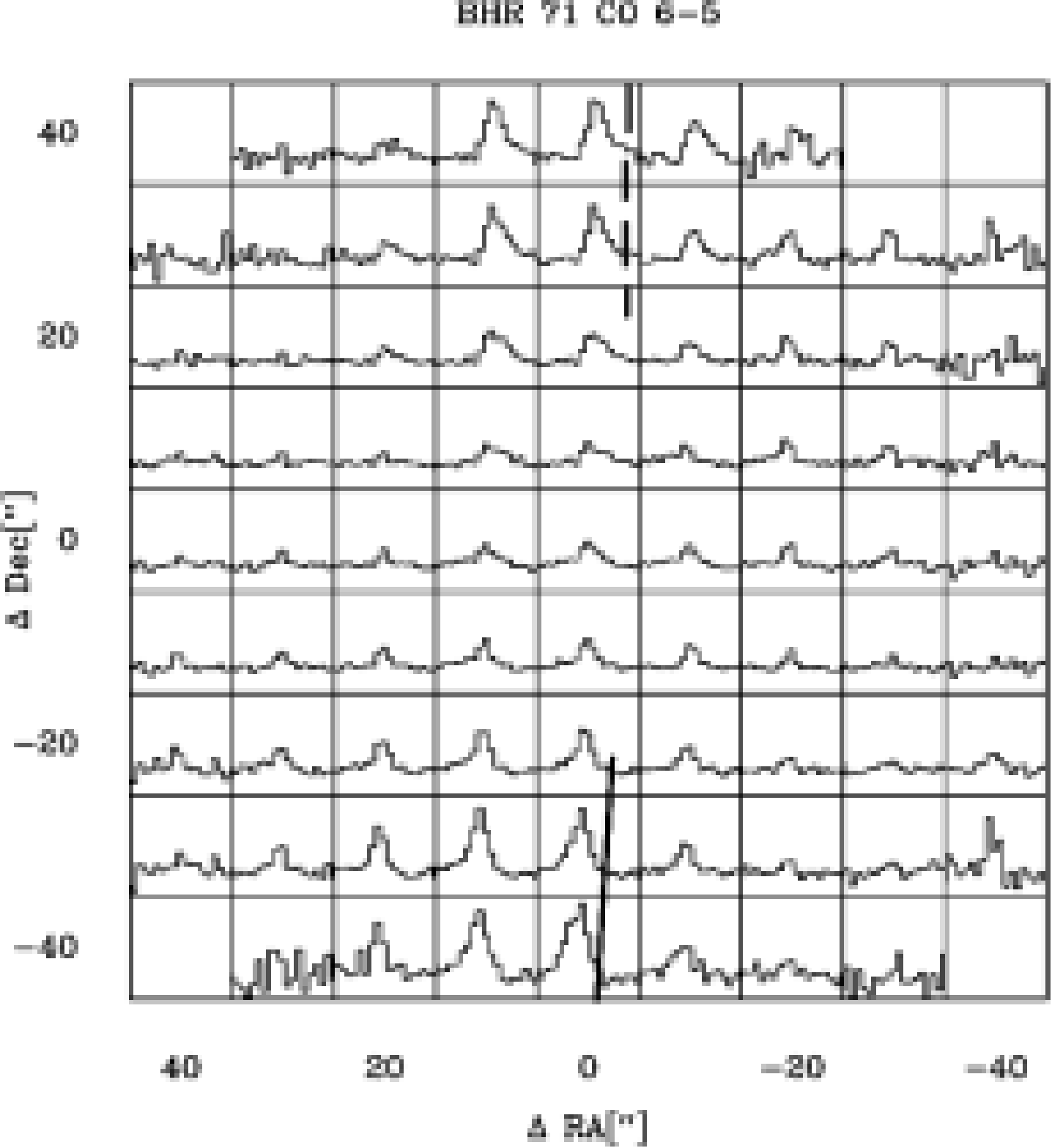}
\end{center}
\caption{Spectra of $^{12}$CO 6--5   over an area of $40''\times40''$ of Ced 110 IRS 4 ({\it left}) and BHR~71 (({\it right}).  For Ced 110 IRS 4, the horizontal axes are -10 to 25 km s$^{-1}$ with vertical axes from -4 to 10 K. For BHR 71, horizontal axes are -20 to 10 km s$^{-1}$ with  vertical axes -5 to 15 K. For BHR~71, the main outflow axis of the red and blue are shown with a {\it dashed} and {\it solid} lines. }
\label{7:fig:spec3}
\end{figure*}
}
\def\placeFigurePaperSevenTwelve{
\begin{figure*}
\begin{center}
\includegraphics[angle=270,width=200pt]{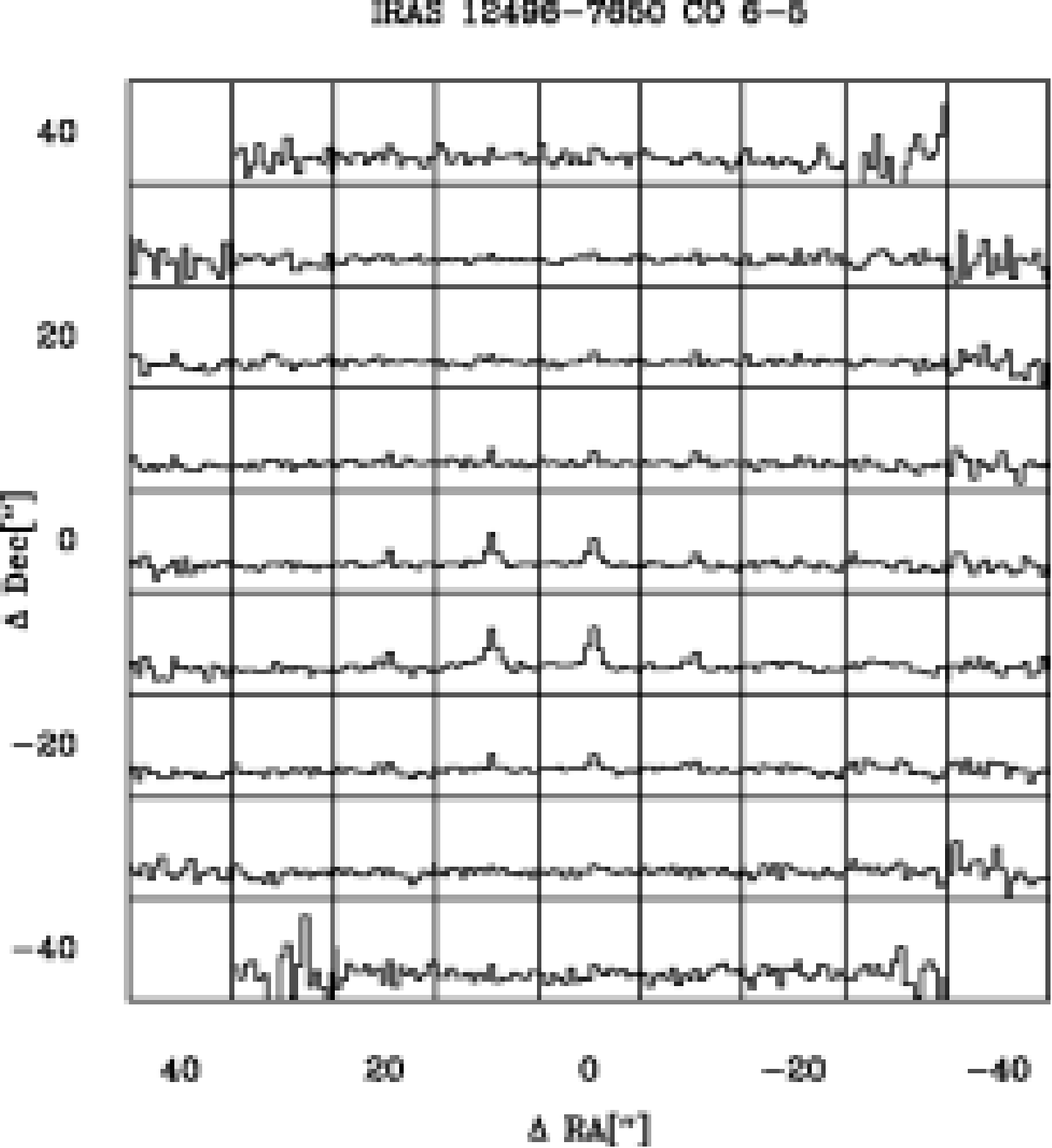}
\includegraphics[angle=270,width=200pt]{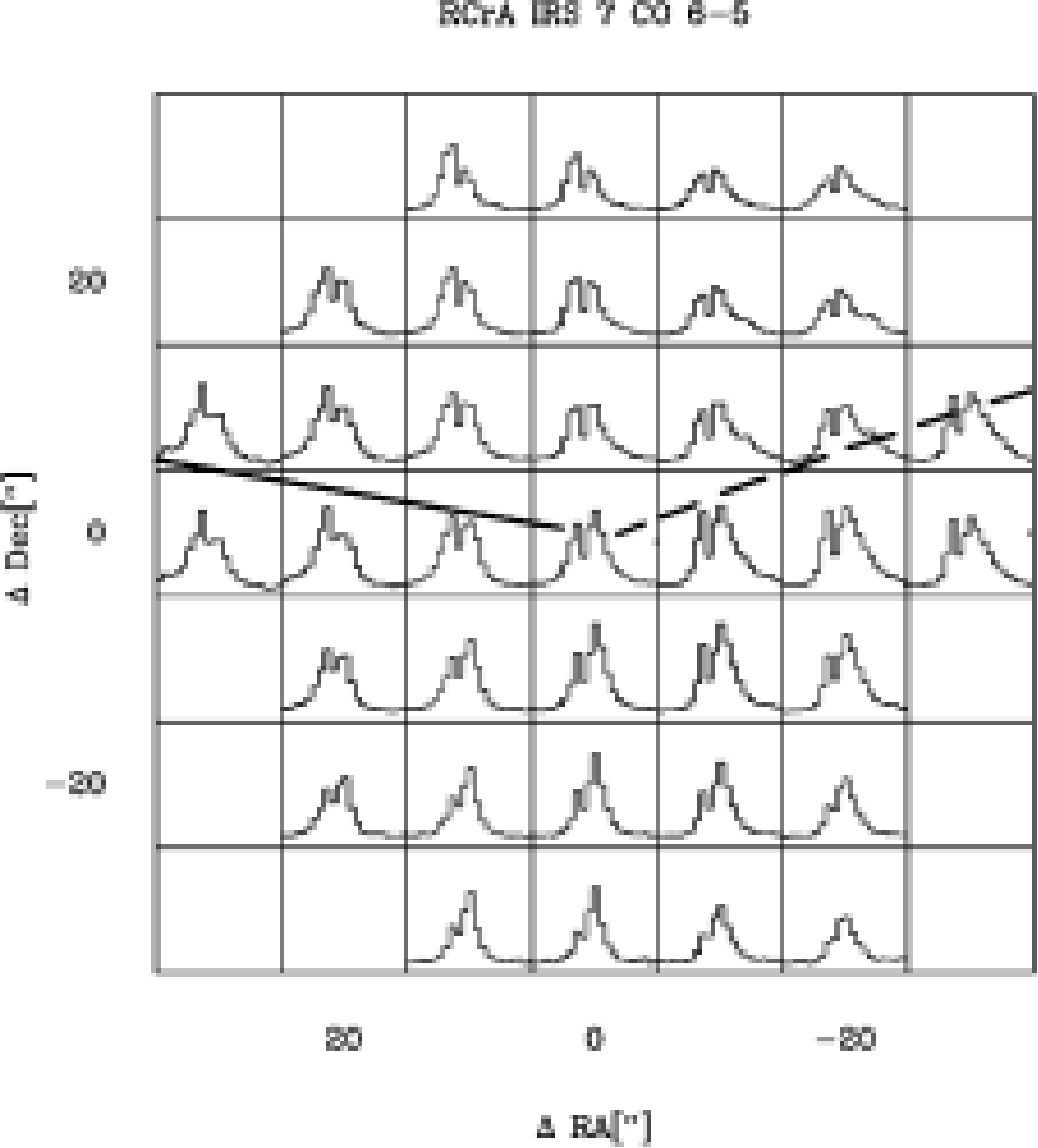}
\end{center}
\caption{Spectra of $^{12}$CO 6--5  over an area of $40''\times40''$ of IRAS 12496-7650 ({\it left}) and RCrA IRS 7A(({\it right}). RCrA IRS 7A was observed using the hexa mode, so the covered area is slightly smaller than that of IRAS 12496-7650 taken in OTF mode. For all spectra, the horizontal axes are -15 to 15 km s$^{-1}$ with vertical axes from -5 to 15 K for IRAS 12496-7650 and from -10 to 25 km s$^{-1}$ and -5 to 50 K for RCrA IRS 7. For RCrA IRS 7, the main outflow axis of the red and blue lobes are shown with a {\it dashed} and {\it solid} lines. }
\label{7:fig:spec4}
\end{figure*}
}

\def\placeFigurePaperSevenThirteen{
\begin{figure*}[th]
\begin{center}
\includegraphics[width=160pt]{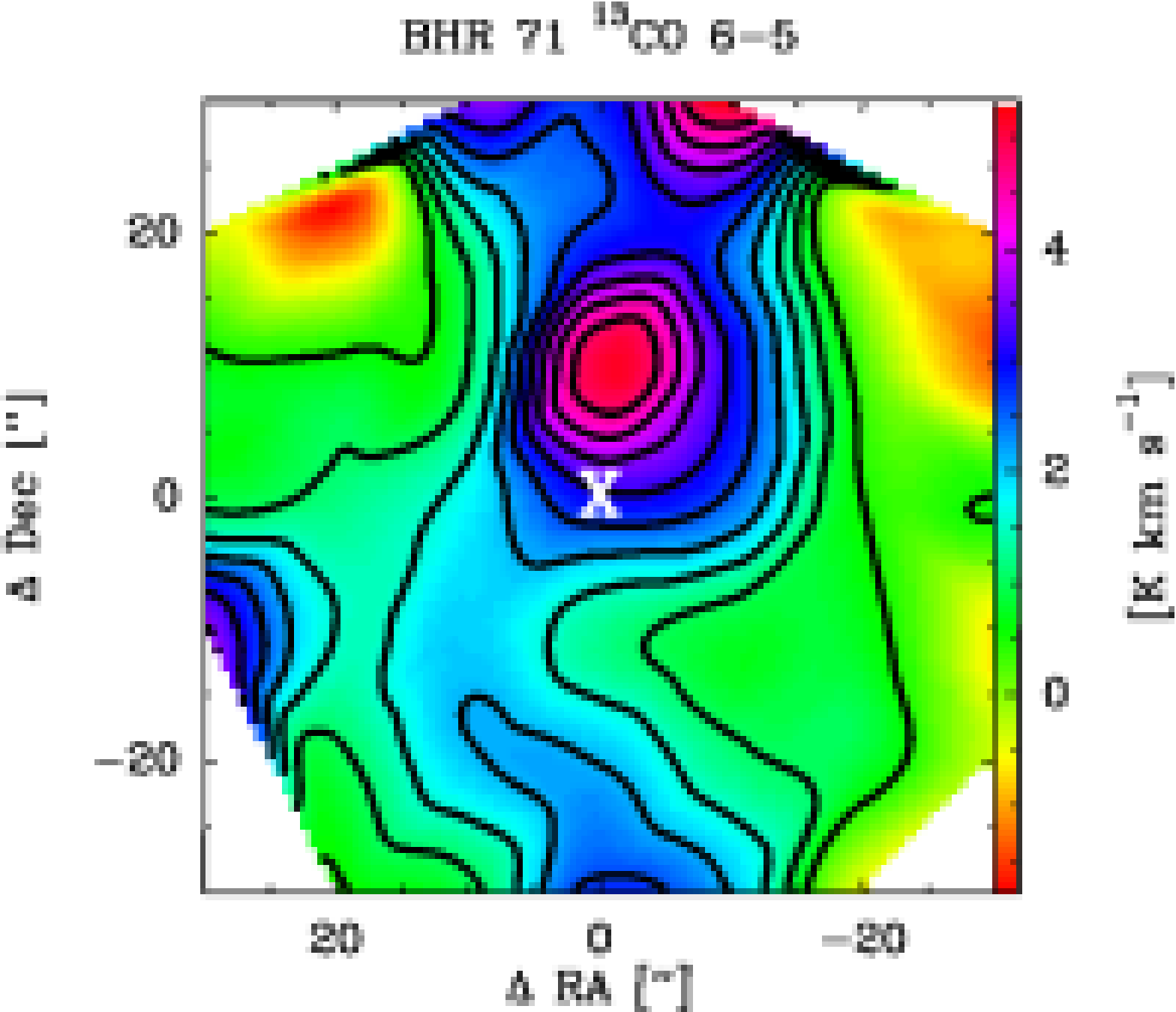}
\includegraphics[width=160pt]{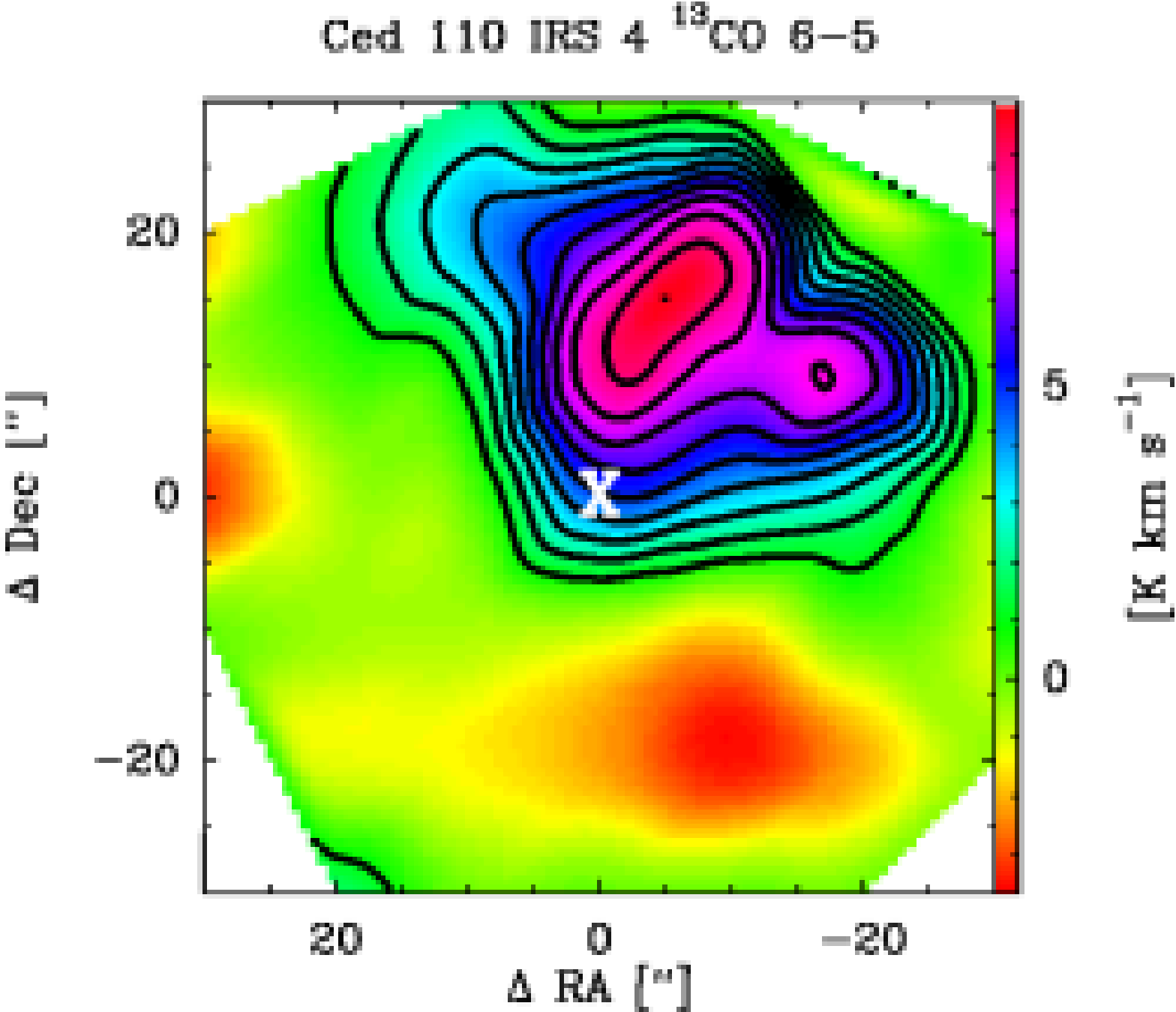}
\includegraphics[width=160pt]{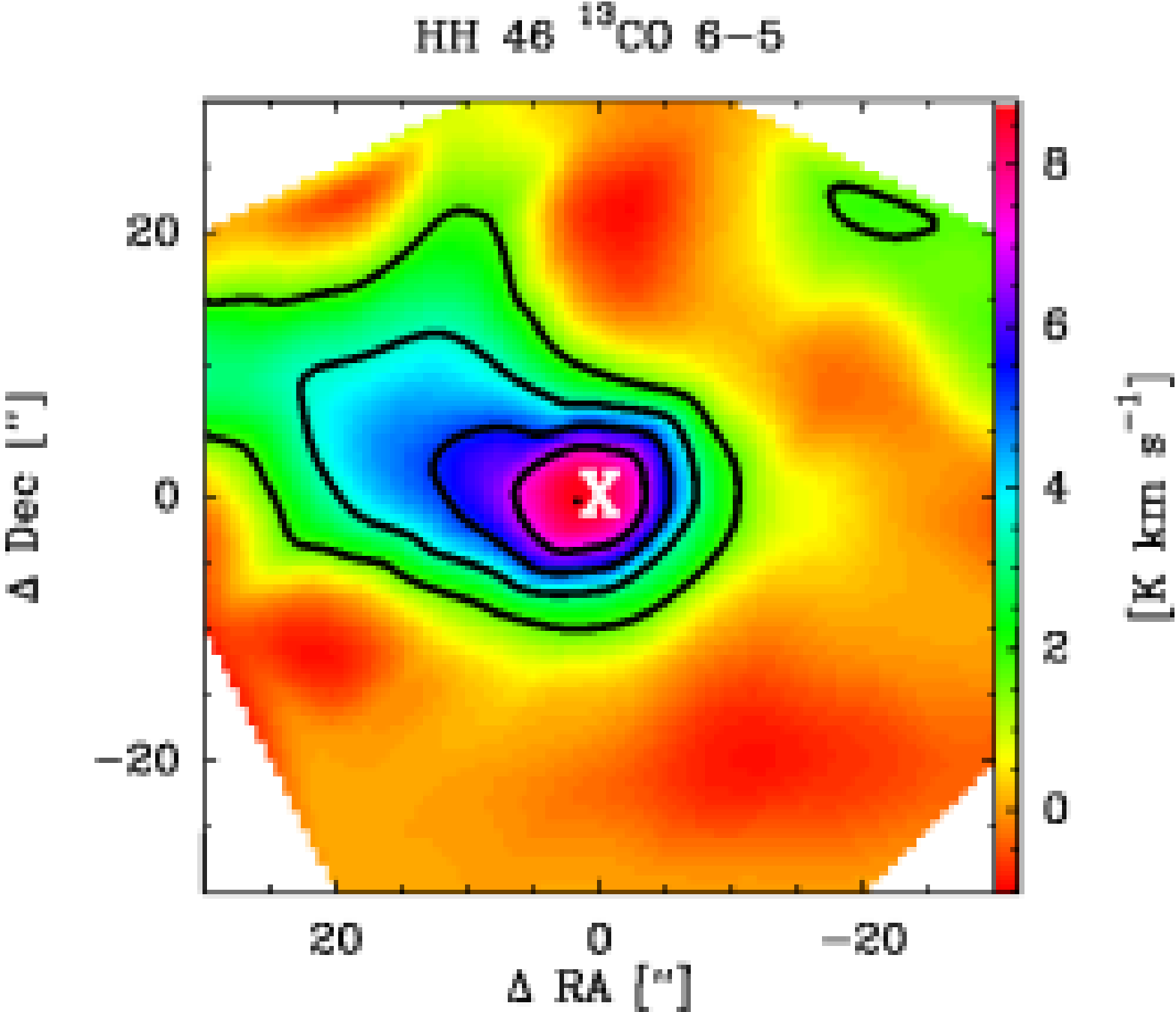}
\includegraphics[width=160pt]{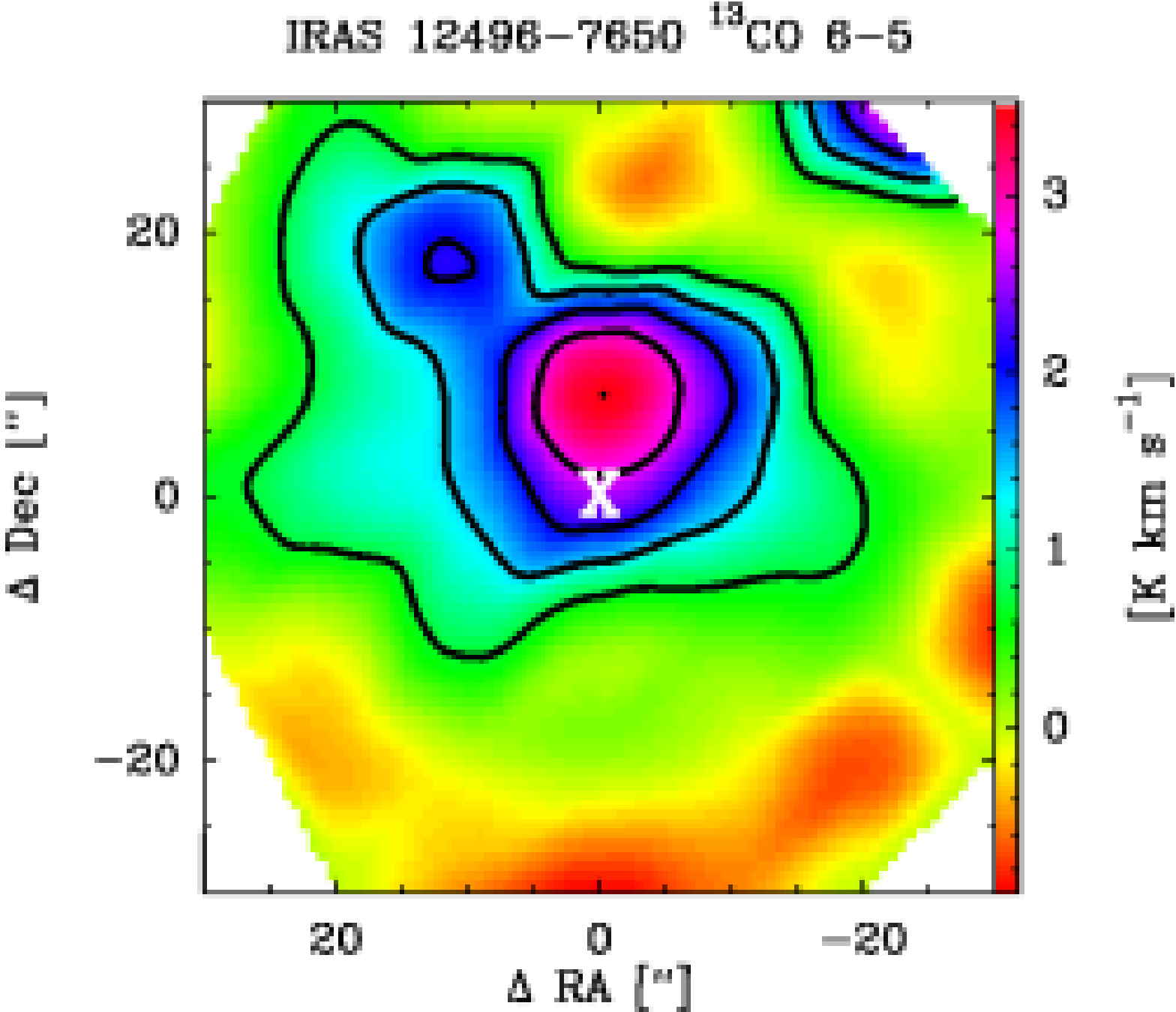}
\includegraphics[width=160pt]{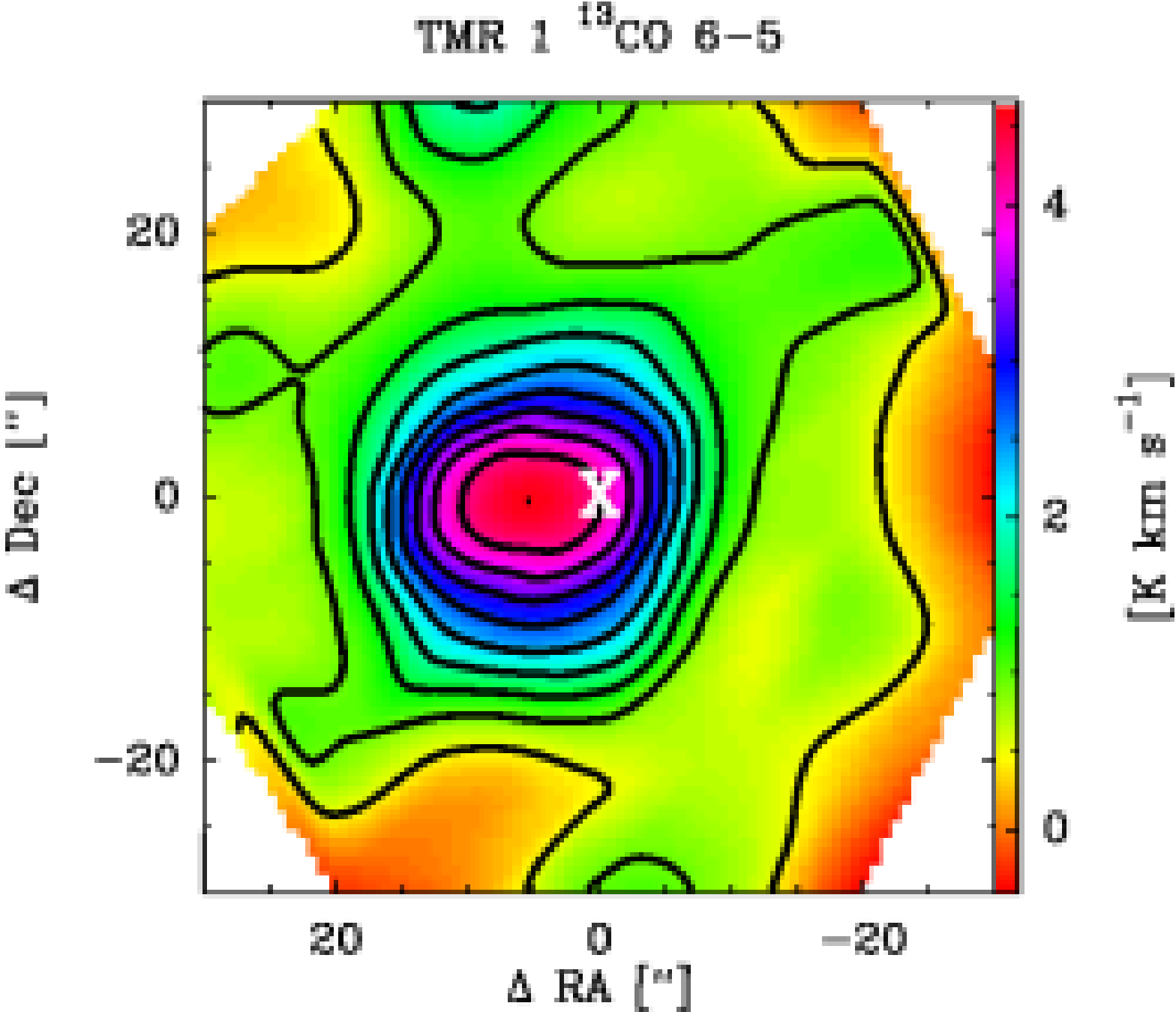}
\end{center}
\caption{$^{13}$CO 6--5 integrated intensity maps of BHR 71 ({\it
    upper left}), Ced 110 IRS 4 ({\it upper middle}), HH~46 ({\it
    upper right}), IRAS 12496-7650 ({\it lower left}) and TMR~1 ({\it
    lower middle}).}  
\label{7:fig:app_1}
\end{figure*}
}
\def\placeFigurePaperSevenFourteen{
\begin{figure*}[th]
\begin{center}
\includegraphics[width=150pt]{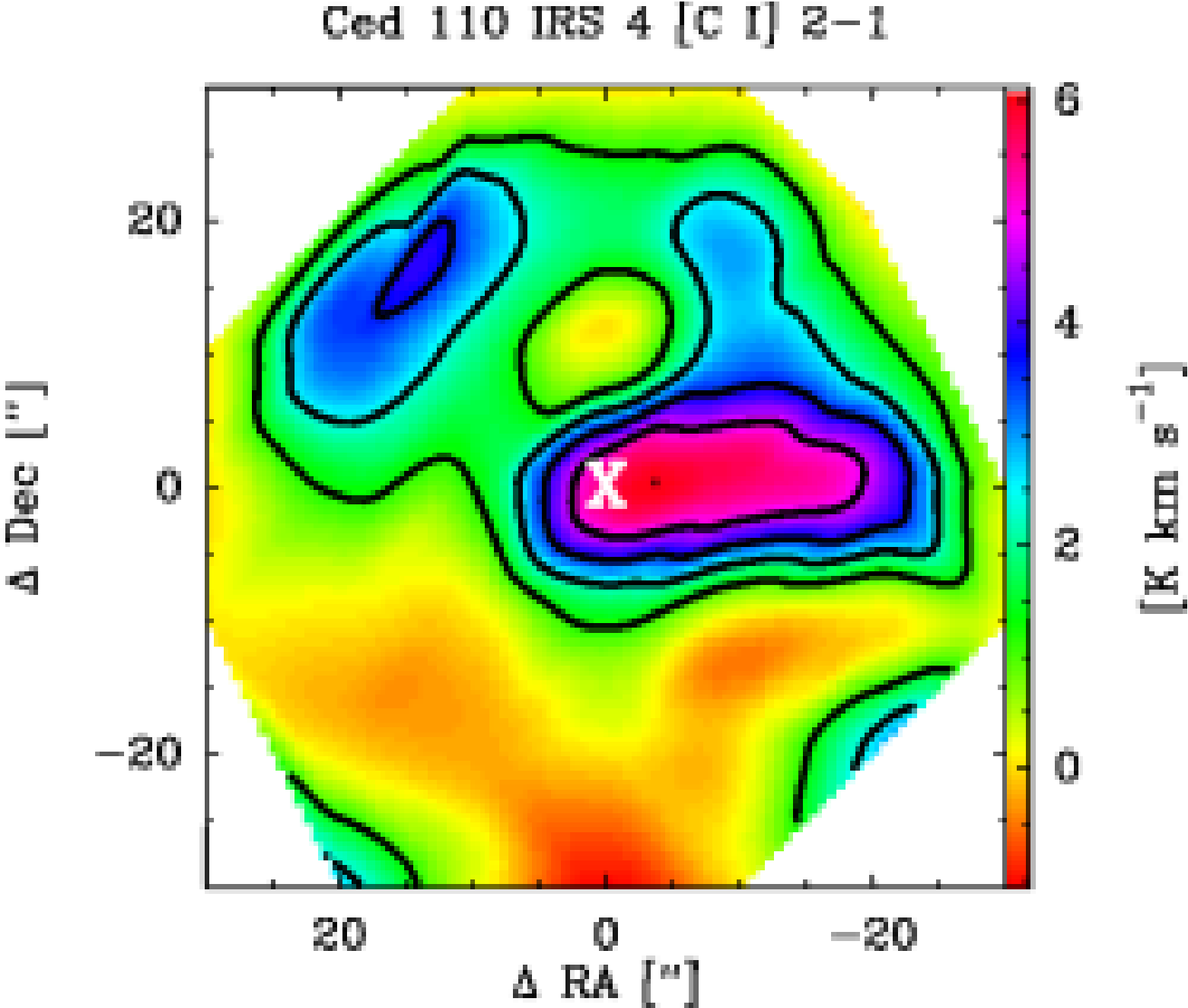}
\includegraphics[width=150pt]{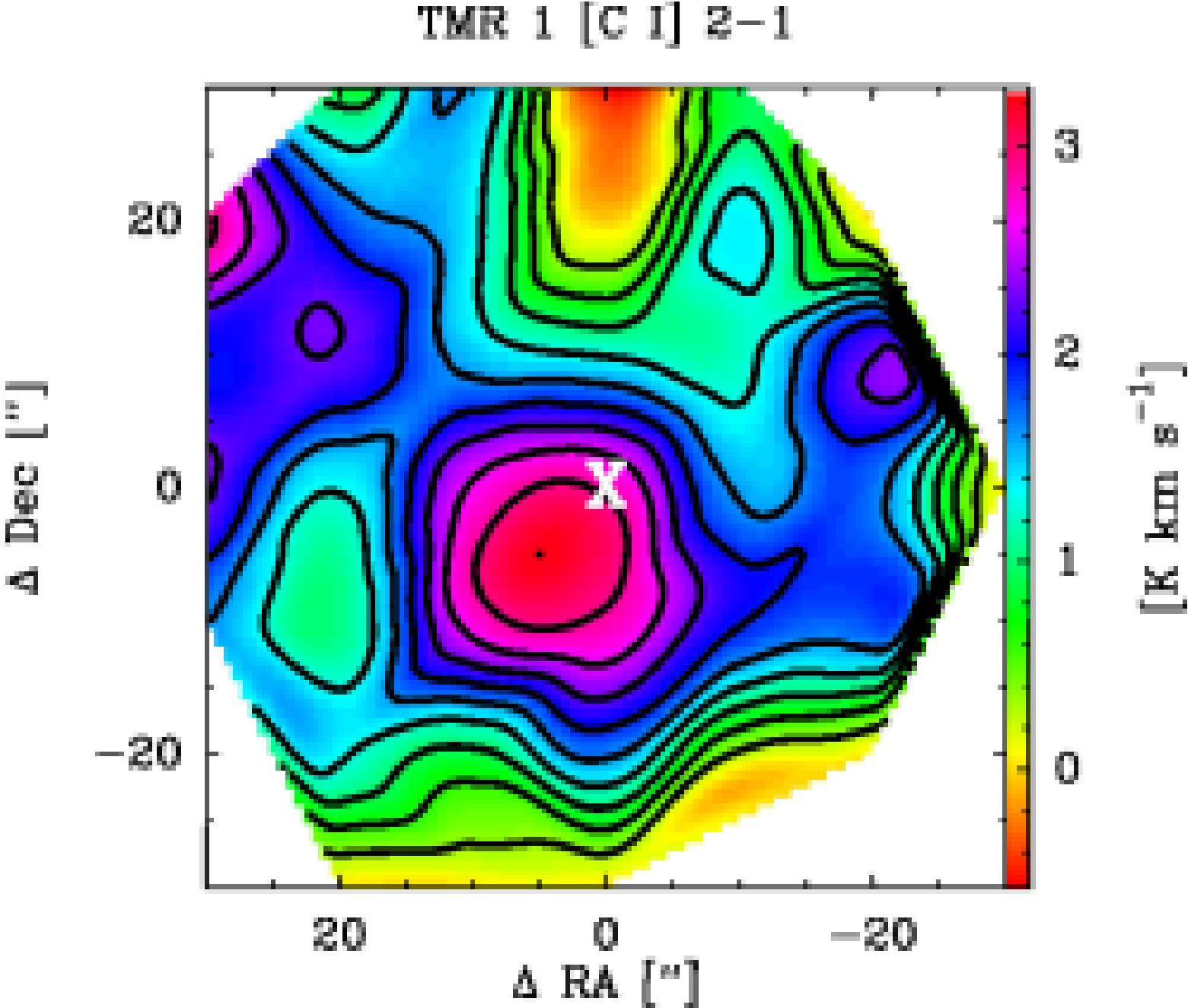}
\end{center}
\caption{[C I] 2--1 integrated intensity maps of Ced 110 IRS 4 ({\it left}) and TMR~1 ({\it right}). Note that the TMR~1 data suffer from a de-rotation problem in the calibrator data and are subject to change.  }
\label{7:fig:app_2}
\end{figure*}
}

\def\placeFigurePaperSevenFifteen{
\begin{figure*}[!th]
\begin{center}
\includegraphics[width=300pt]{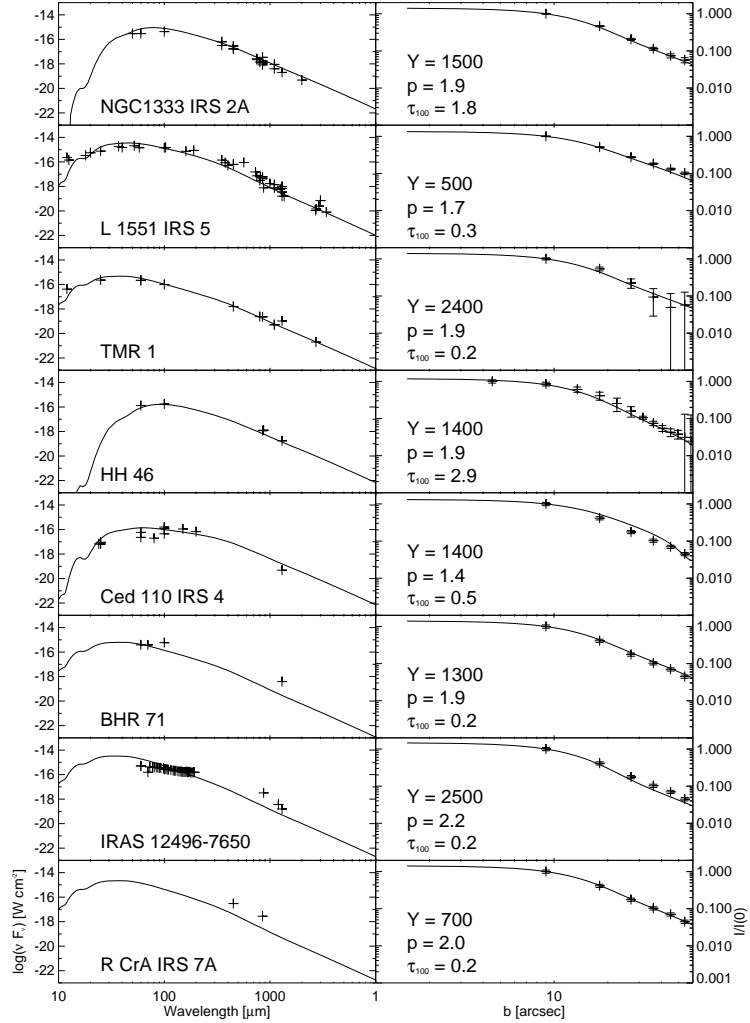}
\end{center}
\caption{The SEDs and radial profiles at 850 $\mu$m of the sample. Overplotted is the best-fitting envelope model. For Ced 110 IRS 4 and BHR 71, the model spatial 850 $\mu$m profile of 1.5 is plotted since data are lacking.  Note that this power law index of 1.5 does not equal the power law index $p$.}
\label{7:fig:dusty}
\end{figure*}
}

\def\placeFigurePaperSevenSixteen{
\begin{figure}[!th]
\begin{center}
\includegraphics[width=250pt]{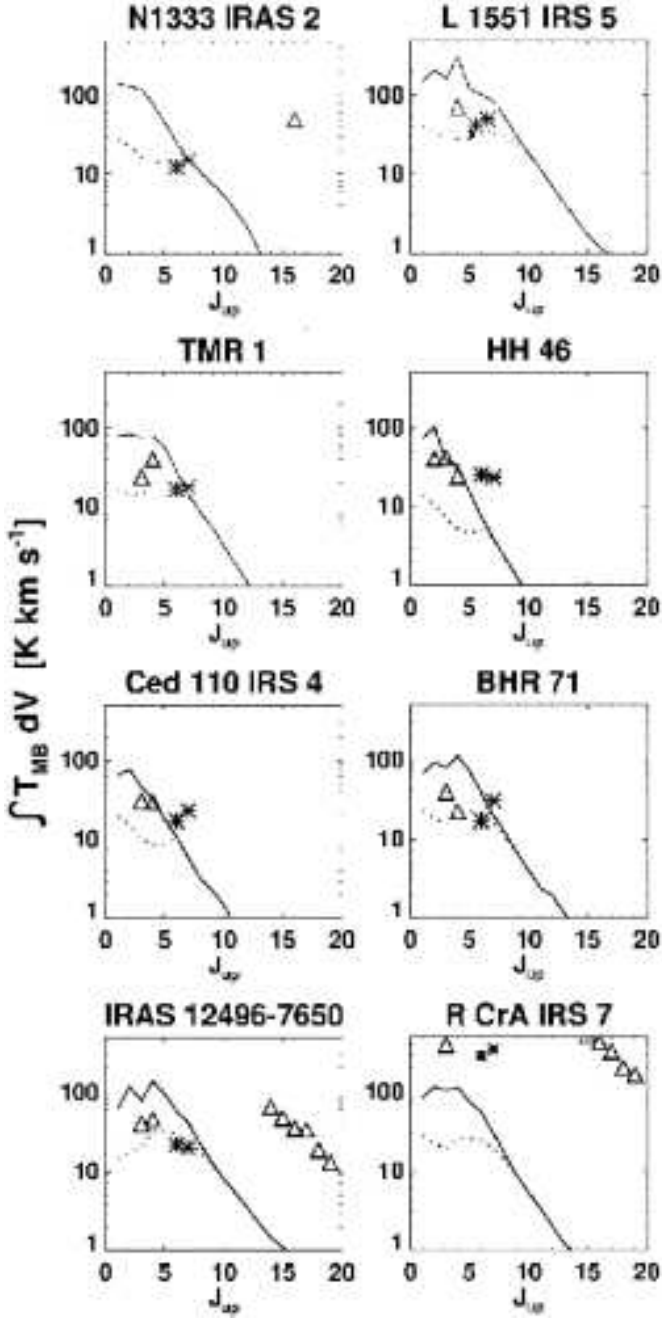}
\end{center}
\caption{The contribution of the envelope to CO lines at the source
  position with $J_{\rm{up}}$ ranging from 1 to 19. Integrated
  emission is shown with dashed and solid lines. Dashed lines are
  modelled lines which show excessive self-absorption. Solid lines
  represent gaussian fitted to the line wings of the envelope model
  and thus represent the strict upper limit to the CO line emission
  from the protostellar envelope. Beam-sizes were taken to be 10$''$
  for all transitions except for $J_{\rm{up}}\leq$3, for which a beam
  of 20$''$ was used. Stars indicate the integrated quiescent emission
  of the lines observed with CHAMP$^+$ from Table \ref{7:tab:COprop}
  at the source position, corrected for red- and blue-shifted
  emission, while triangles are literature data (see Table
  \ref{7:tab:source}).}
\label{7:fig:CO_model}
\end{figure}
}

\def\placeFigurePaperSevenSeventeen{
\begin{figure*}[th]
\begin{center}
\includegraphics[width=170pt]{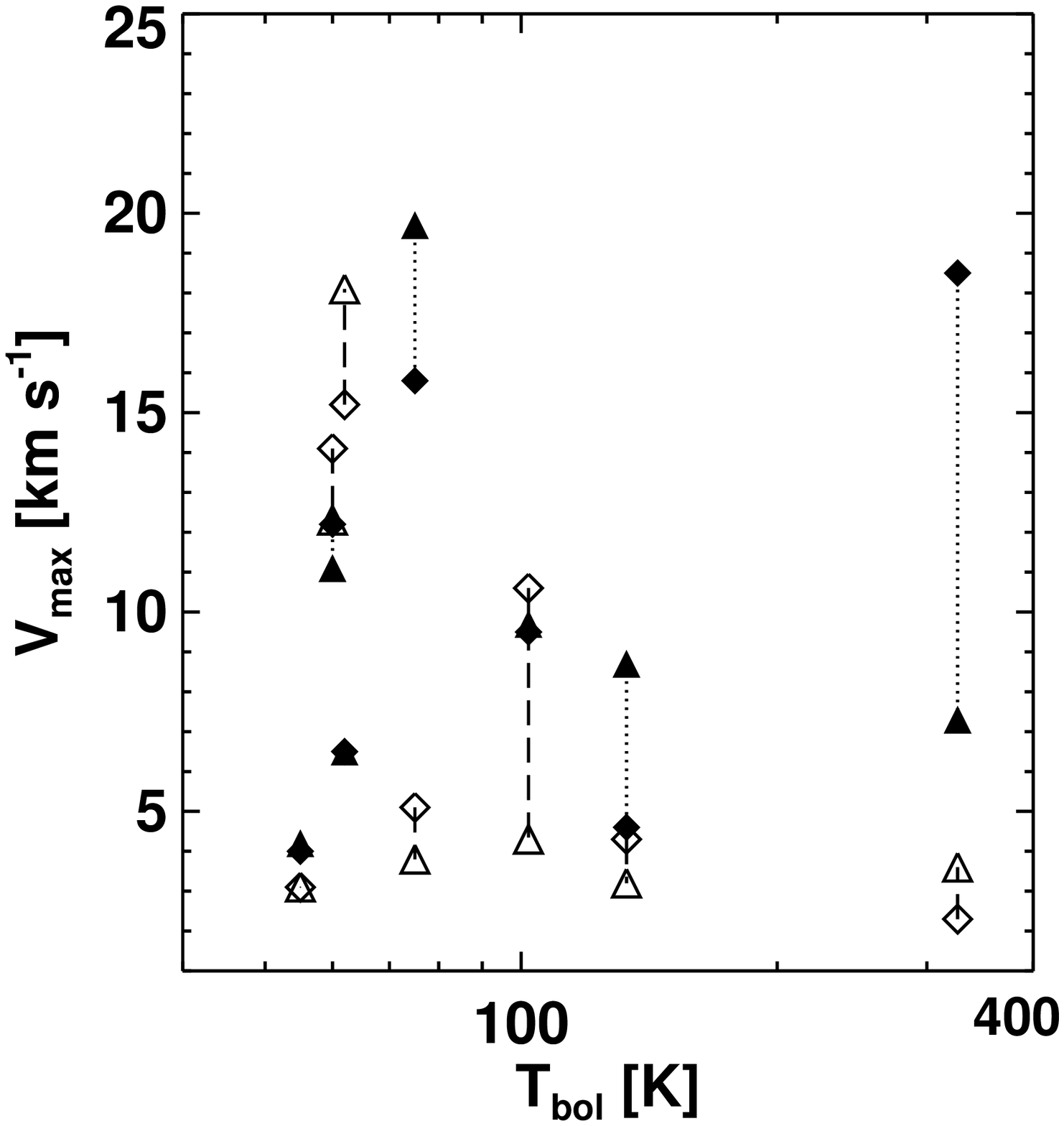}
\includegraphics[width=170pt]{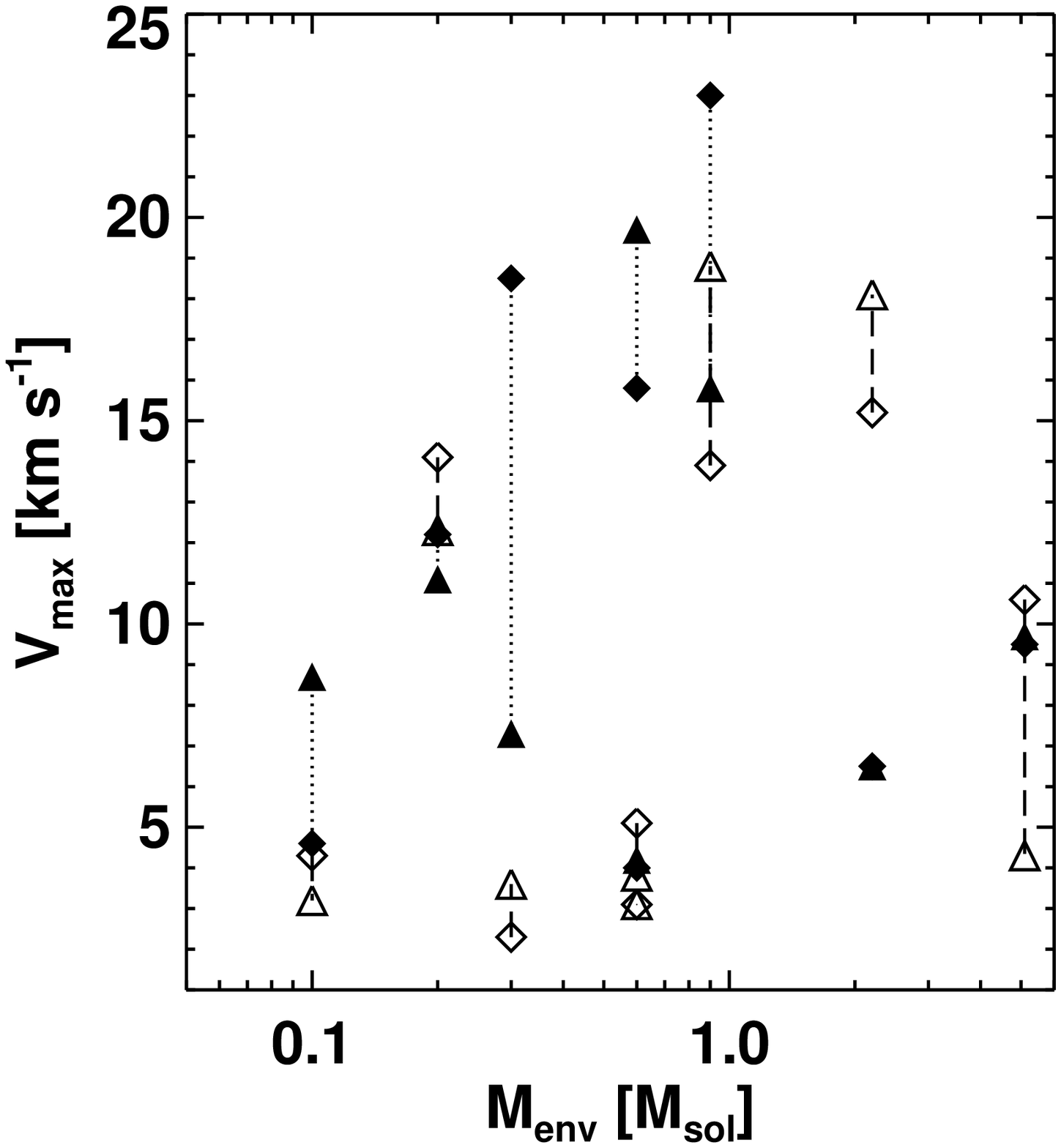}
\includegraphics[width=170pt]{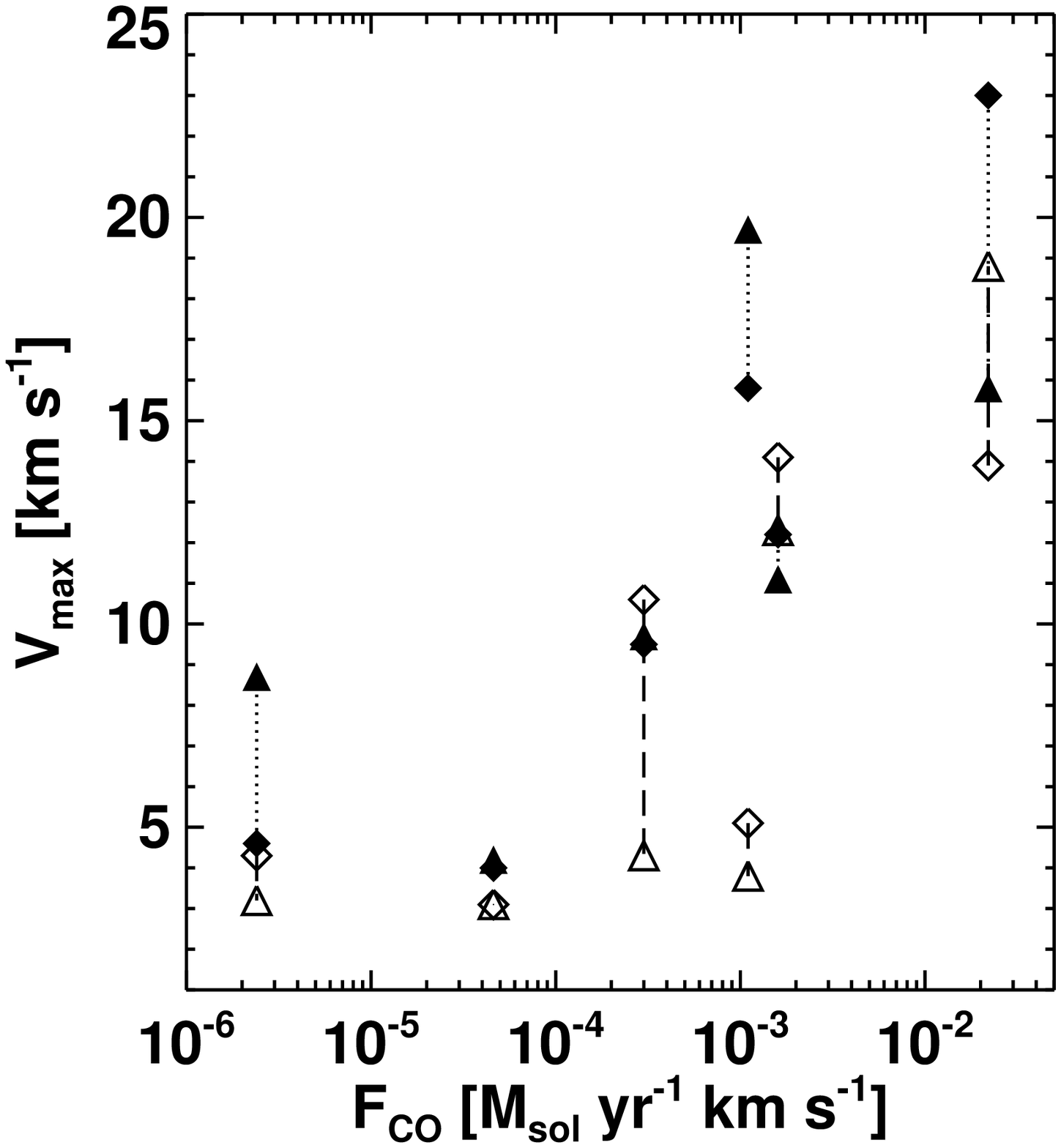}
\end{center}
\caption{$V_{\rm{max}}$ of $^{12}$CO 6--5 and 3--2 vs. $T_{\rm{bol}}$ ({\it left}), envelope mass ({\it middle}) and outflow force.  Triangles represent the blue outflows and diamonds the red outflows. Open symbols are the CO 6--5 lines and filled symbols the CO 3--2 lines. Outflow force from \citet{Cabrit92,Hogerheijde98} (NGC 1333 IRAS 2, L~1551 IRS 5 and TMR 1), \citet{Bourke97} (BHR 71) and van Kempen et al. submitted (Ced 110 IRS 4, IRAS 12496-7650, RCrA IRS 7) and \citet{vanKempen09} (HH~46). CO 3--2 from \citet{Parise06} (BHR 71), \citet{Hogerheijde98} (TMR 1, L 1551 IRS 5), \citet{Knee00} (NGC 1333 IRAS 2), \citet{vanKempen06} (IRAS 12496-7650) and van Kempen et al. 2009, submitted (HH~46, RCrA IRS 7 and Ced 110 IRS 4).}
\label{7:fig:vmax}
\end{figure*}
}

\def\placeFigurePaperSevenEighteen{
\begin{figure*}[th]
\begin{center}
\includegraphics[angle=270,width=500pt]{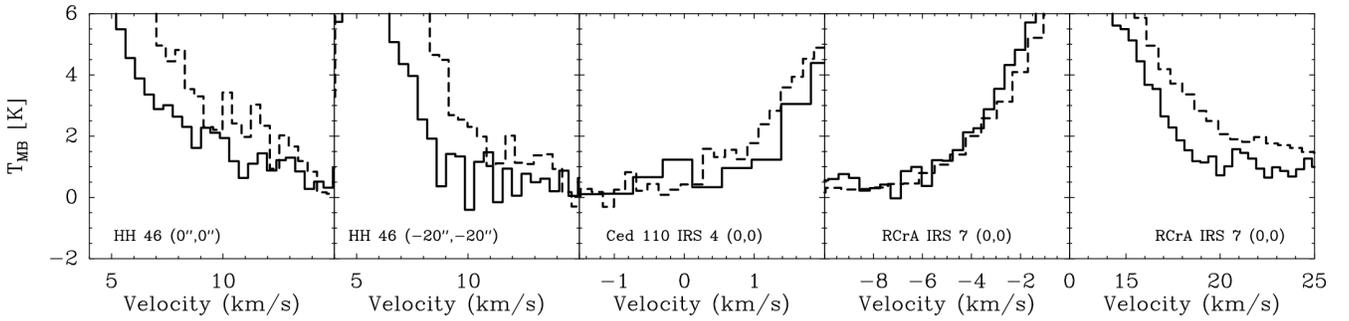}
\end{center}
\caption{Line wings of CO 6--5 ({\it solid}) and CO 3--2 ({\it dashed}) for HH~46 (red wing at central position and in the red outflow), Ced 110 IRS 4 (blue at the central position), RCrA IRS 7A (red and blue at the central position).}
\label{7:fig:overplot}
\end{figure*}
}

\def\placeFigurePaperSevenNineteen{
\begin{figure}[th]
\begin{center}
\includegraphics[width=200pt]{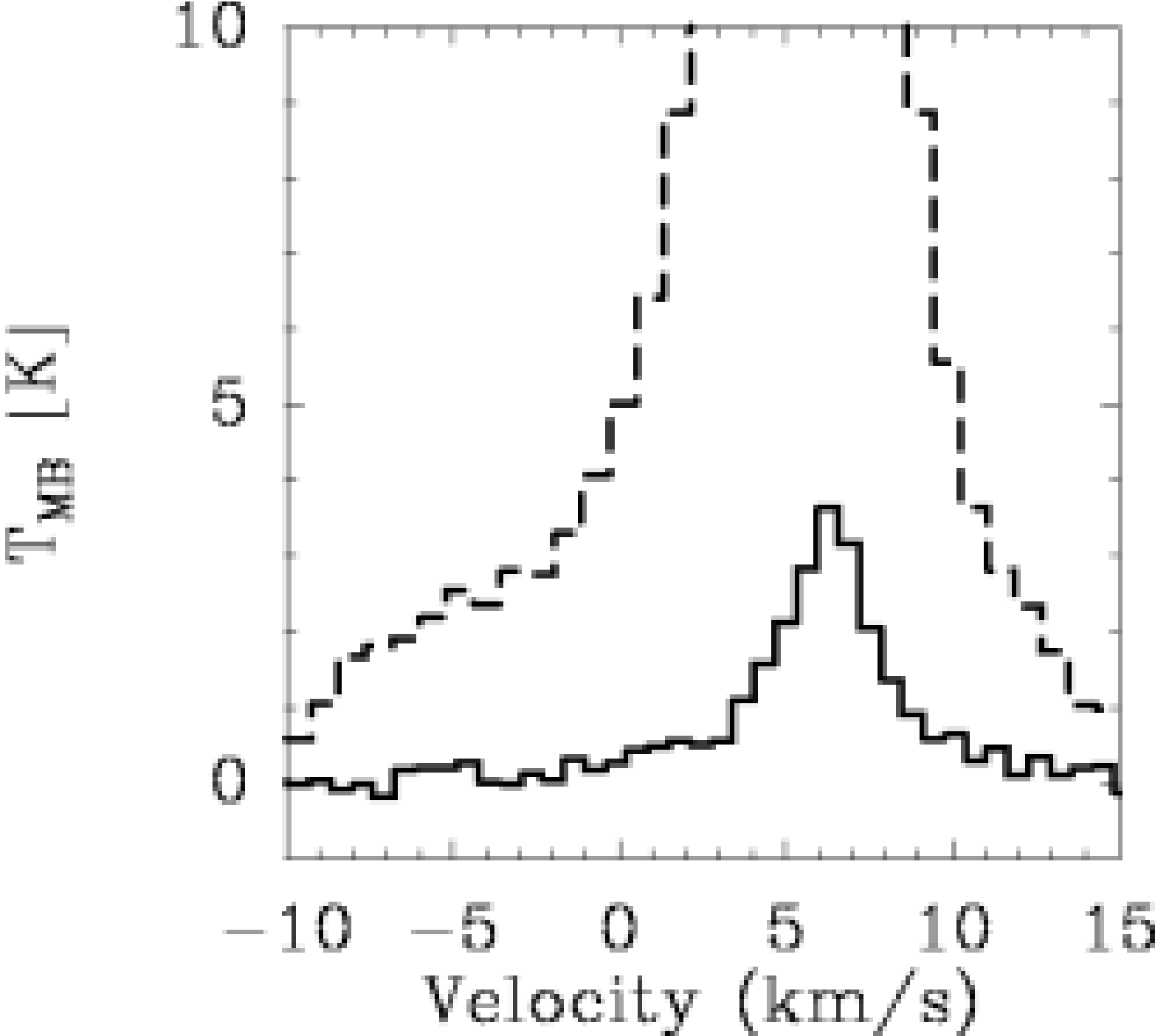}
\end{center}
\caption{CO 4--3 ({\it dashed}) from \citet{Hogerheijde98} overplotted with CO 6--5 ({\it solid}) from the central position of L~1551 IRS 5. The very high CO 4--3/6--5 ratio in both red and blue outflow indicates that the temperatures in this outflow are very low. }
\label{7:fig:L1551spec}
\end{figure}
}



\def\placeTablePaperSevenOne{
\begin{table}
\caption{Adopted Champ$^+$ settings}
\small
\begin{center}
\begin{tabular}{l l l l}
\hline \hline
Setting & Lines & Freq. (GHz) & Mode  \\ \hline
A & $^{12}$CO 6--5 & 691.4730 & OTF/Hexa  \\
A & $^{12}$CO 7--6 & 806.6652 & OTF/Hexa \\
B & $^{13}$CO 6--5 & 661.1067 & Hexa  \\
B & [C I] 2--1    & 809.3435 & Hexa \\
C & C$^{18}$O 6--5 & 658.5533 & Stare \\
C & $^{13}$CO 8--7 & 881.2729 & Stare  \\ \hline
\end{tabular}
\end{center}
\label{7:tab:settings}
\end{table}
}

\def\placeTablePaperSevenTwo{
\begin{table*}
\caption{The sample of sources observed with CHAMP+.}
\small
\begin{center}
\begin{tabular}{l l l l l l l l}  
\hline \hline
Source & RA & Dec & D & $L_{\rm{bol}}$ & $T_{\rm{bol}}$ & Class   \\
 & (J2000) & (J2000) & (pc)& ($L_{\rm{\odot}}$) & (K) &   \\ \hline
NGC 1333 IRAS 2	& 03:28:55.2& +31:14:35  & 250 & 12.7 & 62  & 0  \\    
L1551 IRS 5     & 04:31:34.1&+18:08:05.0 & 160 & 20   & 75  & 1 \\
TMR 1		& 04:39:13.7& +25:53:21  & 140 & 3.1 & 133  & 1 \\  %
HH 46		& 08:25:43.8& -51:00:35.6& 450 & 16 & 102   & 1  \\ 
Ced 110 IRS 4	& 11:06:47.0& -77:22:32.4& 130 & 0.8 & 55   & 1 \\      
BHR 71		& 12:01:36.3& -65:08:44  & 200 & 11 & 60    & 0 \\ 
IRAS 12496-7650	& 12:53:17.2& -77:07:10.6& 250 & 24 &326$^b$& 1  \\        
RCrA IRS 7A$^c$	& 19:01:55.2& -36:57:21.0& 170 & - & -   & 0   \\
\hline 
& Sett.$^{a}$ & \multicolumn{2}{l}{Ref. Cont.} & \multicolumn{3}{l}{Ref.CO} \\ \hline
NGC 1333 IRAS 2	& A   & \multicolumn{2}{l}{1,2,3,4,5,6,7}   & vii \\
L1551 IRS 5     & A   & \multicolumn{2}{l}{2,8,9,10,11,12}  & i \\
TMR 1		& AB  & \multicolumn{2}{l}{13,14,15,16}     & i \\
HH 46		& ABC & \multicolumn{2}{l}{1,17,18}         & ii,iii \\
Ced 110 IRS 4	&  AB & \multicolumn{2}{l}{7,19,20,21,22}   & iii,iv \\
BHR 71		&  AB & \multicolumn{2}{l}{17,23,24,25}     & v \\
IRAS 12496-7650	& AB  & \multicolumn{2}{l}{26, this work}   & iii,viii\\
RCrA IRS 7A$^c$	& A   & \multicolumn{2}{l}{1,7,27,28}       & iii, viii\\ \hline
\end{tabular}
\end{center}
$^a$ Observed settings with CHAMP+. See Table \ref{7:tab:settings} \\
$^b$ IRAS 12496-7650 is  likely to be viewed face-on and thus has strong IR emission and a high $T_{\rm{bol}}$. See van Kempen et al. 2009 submitted\\
$^c$ RCrA IRS 7 is a binary. Pointing was chosen to be on RCrA IRS 7A, believed to be the main embedded source, following results from van Kempen et al. 2009 submitted and \citet{Groppi07}. RCrA IRS 7B is located 15$''$ to the east.  At mid- and far-IR wavelengths, the source is heavily confused with RCrA. Therefore, no reliable $T_{\rm{bol}}$ and $L_{\rm{bol}}$ could be derived. \\
{\it Continuum References: }
1: \citet{diFrancesco08} 
2: \citet{Motte01} 
3: \citet{Gutermuth08}
4: \citet{Enoch06} 
5: \citet{Hatchell07} 
6: \citet{Sandell01}
7: \citet{Froebrich05}
8: \citet{Chandler00}
9: \citet{Reipurth02}
10: \citet{Osorio03}
11: \citet{Liu96}
12: \citet{Butner91}
13: \citet{Kenyon94}
14: \citet{Hogerheijde99}
15: \citet{Furlan08}
16: \citet{Terebey93}
17: \citet{Henning98}
18: \citet{vanKempen09}
19: \citet{Reipurth93}
20: \citet{Luhman08}
21: \citet{Lehtinen01}
22: \citet{Lehtinen03}
23: \citet{Evans07}
24: \citet{Bourke97}
25: \citet{Bourke01}
26: \citet{Henning93}
27: \citet{Nutter05}
28: \citet{Groppi07}\\
{\it CO references : }
i: \citet{Hogerheijde98}
ii: van Kempen et al. 2009 submitted
iii: Paper I
iv: \citet{Hiramatsu07} 
v: \citet{Parise06}
vi: \citet{Knee00}
vii: \citet{Giannini01}
viii: \citet{Giannini99} 
 \\
\label{7:tab:source}
\end{table*}
}

\def\placeTablePaperSevenThree{
\begin{table*}[th]
\caption{Properties of the $^{12}$CO lines$^a$.}
\footnotesize
\begin{center}
\begin{tabular}{l c r r r r r r}  
\hline \hline
Source &   Offpos.  &$V_{LSR}$     & \multicolumn{5}{c}{CO 6--5}            \\
& & & \multicolumn{5}{c}{\hrulefill} \\
       &           &        & $\int T_{\rm{MB}} dV$ & $T_{\rm{peak}}$ &  red  & blue & RMS$^b$ \\ 
       &($''$,$''$)& km s$^{-1}$ & K km s$^{-1}$        & K  & \multicolumn{2}{c}{K km s$^{-1}$}  & K           \\ \hline
NGC 1333 IRAS 2    & 0,0 & 7.0 & 34.3 & 6.8 & 8.9  & 13.2 & 0.6 \\
               & 0,20 & 7.0 & 43.7 & 9.8 & 20.0 & -    & 0.6 \\
            & -20,-20 & 7.0 & 36.3 & 7.6 & 2.2  & 11.0 & 0.7 \\
L 1551 IRS 5    & 0,0 & 6.8 & 39.5 & 7.8 & 13.8 & 10.7 & 0.3 \\
TMR 1           & 0,0 & 4.5 & 23.4 & 8.1 & 2.5  & 5.2  & 0.3 \\ 
HH 46           & 0,0 & 4.4 & 45.3 & 9.3 & 10.6 & 10.4 & 0.3 \\
            & -20,-20 & 4.4 & 33.4 & 9.1 & 11.1 & -    & 0.5\\
Ced 110 IRS4    & 0,0 & 3.5  & 20.8 & 7.8  & -   & 4.8 & 0.4 \\
BHR 71          & 0,0 & -4.5 & 38.6 & 5.2  & 9.1 & 12.7& 0.4 \\
       & 10, -40      & -4.5 & 96.3 & 16.5 & -   & 68.1& 1.0 \\
       & 0, 40        & -4.5 & 75.8 & 13.0 & 41.5& -   & 1.0 \\
IRAS 12496-7650 & 0,0 & 1.8  & 22.1 & 5.9  &  -  & -   & 0.5 \\
RCrA IRS 7A$^c$     & 0,0 & 5.7 & 337.9 & 36.3 & 33.6& 10.9& 0.3 \\ \hline
Source &   Offpos.  &$V_{LSR}$ & \multicolumn{5}{c}{CO 7--6}  \\ 
& & &  \multicolumn{5}{c}{\hrulefill} \\ 
& & & $\int T_{\rm{MB}} dV$ & $T_{\rm{peak}}$ & red  &  blue & RMS$^b$ \\ 
 &($''$,$''$)& km s$^{-1}$  & K km s$^{-1}$        & K             &  \multicolumn{2}{c}{K km s$^{-1}$} & K \\\hline
NGC 1333 IRAS 2    & 0,0 & 7.0 & 29.2 & 6.3 & 8.8 & 5.2 & 1.7 \\
               & 0,20 & 7.0 & 65.8 & 10.1& 27.8& 5.2 & 1.7\\
            & -20,-20 & 7.0 &-    & -    & -   & -   & 3.5\\
L 1551 IRS 5    & 0,0 & 6.5 & 52.9 & 9.2 & 18.9 & 11.6 & 0.4\\
TMR 1           & 0,0 & 4.5 & 19.8 & 10.1& 1.5 & 1.8 & 0.7\\
HH 46           & 0,0 & 4.4 &34.2 & 8.4 & -    & 9.9 & 0.8\\
            & -20,-20 & 4.4 & 29.8 & 6.2 & 15.7 & -   & 1.0\\
Ced 110 IRS4    & 0,0 & 3.5  &23.1 & 7.1 & -    & -   & 1.0\\
BHR 71          & 0,0 & -4.5 &36.6 & 4.4 & 5.4  & -   & 0.7\\
       & 10, -40      & -4.5 &41.5 & 7.0 & 5.6  &17.2 & 2.8\\
       & 0, 40        & -4.5 &54.7 & 9.1 & 27.0 & -   & 3.0\\
IRAS 12496-7650 & 0,0 & 1.8  &19.9 & 6.1 & -    & -   & 1.4\\
RCrA IRS 7A$^c$     & 0,0 & 5.7 &407.6& 46.3& 29.0 & 22.2& 0.7\\ \hline
\hline 
\end{tabular}
\end{center}
\small
$^a$  Red is the redshifted outflow lobe calculated from -10 to -1.5 km s$^{-1}$ with respect to the source velocity. Blue is the blueshifted outflow lobe calculated from +1.5 to +10 km s$^{-1}$ with respect to the source velocity.\\
$^b$ 1$\sigma$ in a 0.8 km s$^{-1}$ channel.\\
$^c$ Outflowing gas refers to emission in  -20 to -8 km s$^{-1}$ and +8 to +20 km s$^{-1}$ from line center.
\label{7:tab:COprop}
\end{table*}
}

\def\placeTablePaperSevenFour{
\begin{table*}[!th]
\caption{CO isotopologue and [C I] properties at the source position.}
\small
\begin{center}
\begin{tabular}{l l l l l l l l l }  
\hline \hline
Source &  \multicolumn{3}{c}{$^{13}$CO 6--5} & $^{13}$CO 8--7 & C$^{18}$O 6--5 & [C I] 2--1 \\ 
 & \multicolumn{3}{c}{\hrulefill}  & & \\
& $\int T_{\rm{MB}} dV$ & $T_{\rm{peak}}$ & FWHM & $\int T_{\rm{MB}} dV$ & $\int T_{\rm{MB}} dV$  & $\int T_{\rm{MB}} dV$  \\
& K km s$^{-1}$ & K & km s$^{-1}$ &  K km s$^{-1}$ & K km s$^{-1}$ & K km s$^{-1}$ \\ \hline
TMR 1       & 4.3 & 1.9  & 2.0 & - & - & 2.9\\
HH 46           & 3.9 & 3.1 & 1.5 &$<$0.3$^a$ & $<$0.15$^a$ & 2.3 \\
Ced 110 IRS4    & 5.2 & 2.1 & 2.3& - & - & 2.5\\
BHR 71          & 1.4 & 1.5  & 0.8 & - & - & $<$0.6$^a$\\
IRAS 12496-7650 & 3.8 & 1.7  & 2.1& - &  - & $<$0.6$^a$\\
\hline 
\end{tabular}\\
\end{center}
\label{7:tab:isoprop}
$^a$ 1$\sigma$ level in a 0.7 km s$^{-1}$ bin.\\
\end{table*}
}

\def\placeTablePaperSevenFive{
\begin{table*}
\caption{Inferred envelope properties from the DUSTY models. }
\begin{center}
\small
\begin{tabular}{l r r r r r r r}
\hline \hline  
Source  & Y & $p$ & $\tau_{100}$ & $R_{\rm{inner}}$ & $R_{\rm{outer}}$ & $n_{\rm{1000 AU}}$ & $M_{\rm{env}}$($<R_{\rm{outer}}$) \\
 & & & & (AU) & (10$^4$ AU) & (10$^4$ cm$^{-3}$) & (M$_{\odot}$) \\ \hline
RCrA IRS 7A     & 700  & 2.0 & 0.2 & 24.7 & 5.9 & 13 & 0.9 \\
NGC 1333 IRAS 2 & 1500 & 1.9 & 1.8 & 31.3 & 1.6 & 140& 2.2 \\ 
L~1551 IRS 5    & 500  & 1.7 & 0.3 & 26.5 & 1.3 & 33 & 0.6 \\
TMR 1           & 2400 & 1.9 & 0.2 & 9.5  & 2.3 & 5.4 &0.1 \\
IRAS 12496-7650 & 2500 & 2.2 & 0.2 & 35.7 & 8.9 & 8.7 &0.3\\
HH 46           & 700  & 1.8 & 2.6 & 30.0 & 2.1 & 250 &5.1  \\ 
BHR 71          & 1300 & 1.9 & 0.2 & 15.6 & 2.0 & 8.5 &0.2\\
Ced 110 IRS 4   & 1400 & 1.4 & 0.5 & 5    & 0.7 & 51 & 0.6\\ \hline
\end{tabular}\\
\end{center}
\label{7:tab:env}
\end{table*}
}

\def\placeTablePaperSevenSix{
\begin{table}
\caption{$^{13}$CO model predictions}
\begin{center}
\small
\begin{tabular}{l l l l l}
\hline \hline  
Source  & \multicolumn{2}{c}{$\int T_{\rm{MB}}dV$} & \multicolumn{2}{c}{$T_{\rm{peak}}$} \\ 
& \multicolumn{2}{c}{\hrulefill} &\multicolumn{2}{c}{\hrulefill}\\
& Mod. & Obs. & Mod. & Obs. \\ \hline
TMR 1           & 3.6 &  4.3 &1.9 &1.9  \\
IRAS 12496-7650 & 7.2 &  3.3 &3.6 & 1.6 \\
HH 46           & 1.9 &  3.9 &1.2 & 2.9 \\
BHR 71          & 4.7 &  1.4 &2.7 & 1.2 \\
Ced 110 IRS 4   & 2.3 &  5.2$^a$ &1.2 & 1.9\\ \hline
\end{tabular}\\
\end{center}
$^a$ Possible outflow contribution, see Fig. \ref{7:fig:CED110} and \ref{7:fig:app_1}, estimated to be at 25$\%$ of total integrated intensity.
\label{7:tab:13CO}
\end{table}
}

\def\placeTablePaperSevenSeven{
\begin{table}
\caption{Outflow kinetic temperature estimates$^a$. }
\begin{center}
\small
\begin{tabular}{l l l l l}
\hline \hline  
Source  & Ratio$^b$ & $T_{\rm{kin}}$(Blue) & Ratio$^b$ & $T_{\rm{kin}}$(Red)  \\
  & K & K  \\ \hline
RCrA IRS 7      & 1 & $>$200 & 2.5 & 160 \\
NGC 1333 IRAS 2    & - & - & - & -\\
L~1551 IRS 5$^c$& $>$8 & $<$50 & 9 & 40\\
TMR 1$^d$       & 1.5 & 180 & $>$4 & 120 \\
IRAS 12496-7650 & $>$8 & $<$65 & - & -\\
HH 46$^e$       & 6 & 100 & 3 & 140 \\
BHR 71          & - & - & - & -\\
Ced 110 IRS 4   & 3 & 140 & $>$7 & $<$80 \\ \hline
\end{tabular}\\
\end{center}
$^a$ Estimated with an ambient density of 10$^{4}$ cm$^{-3}$. 
\\
$^b$ CO 3--2/6--5 line wing ratio. Ratios are averaged over velocity and distance from the source. See \citet{vanKempen09} for more discussion.\\
$^c$ CO 4--3/6--5 ratio.\\
$^d$ Optical depth of 8 inferred for $^{12}$CO 3--2 wings \citep{Hogerheijde98}.\\
$^e$ See \citet{vanKempen09}. \\
\label{7:tab:temp}
\end{table}
}

\def\placeTablePaperSevenOnlineOne{
\begin{table}[!ht]
\caption{Extracted fluxes from LABOCA}
\begin{center}
\begin{tabular} {l l}
\hline \hline
Source & Flux \\
& Jy (120$''$) \\ \hline
IRAS 12496-7650 & 9.4 \\
HH 52-54 & $<$0.065$^a$ \\
IRAS 12553-7651 & 4.2 \\ \hline 
\end{tabular}\\
\end{center}
$^a$ 3$\sigma$ level in Jy/beam. \\
\label{7:tab:laboc}
\end{table}
}



\abstract{ The origin and heating mechanisms of warm (50$<T<$200~K)
  molecular gas in low-mass young stellar objects (YSOs) are strongly
  debated.  Both passive heating of the inner collapsing envelope by
  the protostellar luminosity as well as active heating by shocks and
  by UV associated with the outflows or accretion have been proposed.
Most data so far have focussed on the colder gas component.}
{We aim to characterize the warm gas within protosteller objects, and
  disentangle contributions from the (inner) envelope, bipolar
  outflows and the quiescent cloud.}  {High-$J$ CO maps ($^{12}$CO
  $J$=6--5 and 7--6) of the immediate surroundings (up to 10,000 AU)
  of eight low-mass YSOs are obtained with the CHAMP$^+$ 650/850 GHz
  array receiver mounted on the APEX telescope.  In addition,
  isotopologue observations of the $^{13}$CO $J=$6--5 transition and
  [C I] $^3P_2-^3P_1$ line were taken.}  {Strong quiescent narrow-line
  $^{12}$CO 6--5 and 7--6 emission is seen toward all protostars. In
  the case of HH~46 and Ced 110 IRS 4, the on-source emission
  originates in material heated by UV photons scattered in the outflow
  cavity and not just by passive heating in the inner envelope. Warm
  quiescent gas is also present along the outflows, heated by UV
  photons from shocks. This is clearly evident in BHR~71 for which
  quiescent emission becomes stronger at more distant outflow
  positions. Shock-heated warm gas is only detected for Class 0 flows
  and the more massive Class I sources such as HH~46. Outflow
  temperatures, estimated from the CO 6--5 and 3--2 line wings, are
  $\sim$100 K, close to model predictions, with the exception of the
  L~1551 IRS 5 and IRAS 12496-7650, for which temperatures $<$50 K are
  found.
} {APEX-CHAMP$^+$ is uniquely suited to directly probe the protostar's
  feedback on its accreting envelope gas in terms of heating,
  photodissociation, and outflow dispersal by mapping $\sim1'\times1'$
  regions in high-$J$ CO and [C I] lines. Photon-heating of the surrounding gas
  may prevent further collapse and limit stellar growth.}


   \maketitle

\section{Introduction}
Low-mass ($M<3$ M$_{\odot}$) Stage 0 and Stage 1 protostars
\citep{Robitaille06} have two distinct components. First, they are
surrounded by a protostellar envelope, consisting of a reservoir of
gas and dust, which can feed the star and disk. The envelope can be
divided into cold ($T\sim10-20$ K) outer regions and a warm ($T>50$ K)
inner region
\citep{Adams87,Shirley00,Schoeier02,Shirley02,Jorgensen02, Whitney03b,
  Robitaille06}.  Second, most, if not all, embedded protostars have
molecular outflows \citep{Bachiller96,Bachiller99,Arce05,Lee05,Hirano06}. These outflows have been
observed to have a wide variety in intensity, collimation and affected
area, ranging from the strong and large (ten thousands of AU) outflows
seen in L~1448, L~1551 IRS 5 and HH~46
\citep[e.g.,][ from here on referred to as Paper I]{Bachiller94,Heathcote96, vanKempen09} to the weak, much
more compact, outflows as found for TMR~1 and TMC-1
\citep{Cabrit92,Hogerheijde98}. Such molecular outflows are observed
to widen with time and influence the evolution of the envelope
\citep[e.g.,][]{Arce05}.

Envelope models can be constructed using either a 1-D or 2-D
self-consistent dust radiative transfer calculation, constrained by
observations of the cold dust \citep[e.g.][]{Shirley02,Jorgensen02}
and/or the Spectral Energy Distribution (SED)
\citep[e.g.][]{Whitney03a,Whitney03b, Robitaille06,
  Crapsi08}. Combining heterodyne observations with radiative transfer
calculations, one can in turn analyze line emission of a large range
of molecules and determine their origin and excitation conditions,
both in the inner and outer regions of protostellar envelopes as well
as in circumstellar disks and fore-ground components
\citep[e.g.][]{Boogert02,Jorgensen04,Maret04,Brinch07,Lee07}. Envelope
models predict that the inner regions are warm ($T>100$ K), dense
($n$(H$_2$) $>10^6$ cm$^{-3}$) and relatively small ($R<$ 500 AU).
Most molecular lines observable at (sub)mm wavelengths trace the
colder gas, but a few lines, such as the high-$J$ CO lines, directly
probe the warm gas. The atmospheric windows at 650 and 850 GHz are the
highest frequency windows in which observations of CO (up to energy
levels of $\sim$150 K) can be routinely carried out. Unfortunately,
few studies have succesfully observed CO transitions or its
isotopologues in these atmospheric windows due to the excellent
weather conditions necessary. In addition, such studies are often
limited to single spectra of only a handful of YSOs
\citep[e.g.][]{Schuster93,Schuster95,Hogerheijde98,Ceccarelli02,Stark04,
  Parise06,vanKempen06}. The lack of spatial information on the warm
gas distribution has prevented an in-depth analysis. Comparison
between ground-based high-$J$ CO observations and far-IR CO
transitions (CO 15--14 and higher) do not always agree on the origin
of the high-$J$ CO lines \citep{vanKempen06}.

Complex molecules such as H$_2$CO and CH$_3$OH, emit in high
excitation lines at longer wavelengths, and have been observed to have
surprisingly high abundances in low-mass protostars
\citep[e.g.][]{vanDishoeck95,Blake95, Ceccarelli00,Schoeier02,
  Maret04,Jorgensen04,Bottinelli04,Bottinelli07}. Unfortunately, the
abundances of these molecules are influenced by the gas-phase and
grain surface chemistry, complicating their use as tracers of the
physical structure \citep[e.g.,][]{Bisschop07}.  It has been proposed
that the emission of such molecules originates inside a hot core
region, a chemically active area close to the star, coinciding with
the passively heated warm inner region of the protostellar envelope
where ices have evaporated
\citep{Ceccarelli00,Bottinelli04,Bottinelli07}. However, to fully
understand the origin of these complex organics, knowledge of the
structure of the warm gas is an essential ingredient.

Warm gas near protostars can have different origins than passive
heating alone. Outflow shocks passing through the envelope can be a
source of heat. Quiescent gas, heated by X-rays or UV, is present in
the inner envelope. For example, \citet{Stauber04} show that
significant amounts of Far Ultra-Violet (FUV) or X-ray photons are
necessary to reproduce the line intensities and the derived abundances
of molecules, such as CO$^+$, CN, CH$^+$ and NO in both high and
low-mass protostars.  \citet{Spaans95} investigated photon heating of
outflow cavity walls to reproduce the observed line intensties and
widths of $^{12}$CO 6--5 and $^{13}$CO 6--5 emission in Class I
sources \citep{Hogerheijde98}.  In paper I, an extension
of this model is proposed in which a Photon Dominated Region (PDR) at
the outflow/envelope cavity walls is present in the HH~46 outflow to
explain the relatively strong and quiescent high-$J$ CO emission. The
emission of [C I] can constrain the color and extent of the more
energetic photons. Apart from the accretion disk, UV photons are also
produced by the jet shocks in the outflow cavity and the outflow bow
shock.  All these models provide different predictions for the spatial
extent of the warm gas ($\sim$1$'$ for the outflow to $<$1$''$ for a
passively heated envelope), as well as the different integrated
intensities and line profiles of the high-$J$ CO transitions.

The most direct tracers of the warm ($50<T<200$ K) gas are thus the
high-$J$ CO lines. So far, studies to directly detect the warm gas
components through these lines have rarely been able to disentangle
the envelope and outflow contributions.  Far Infrared (IR) transitions
of even higher-$J$ CO, the far-IR CO lines, have been observed using
the ISO-LWS instrument to trace the inner regions
\citep[e.g.,][]{Ceccarelli98,Giannini99, Nisini99,Giannini01,
  Nisini02}, but could not unambigiously constrain the origin of the
warm gas emission, due to the limited spatial and spectral resolution.

The Chajnantor plateau in northern Chile, where the recently
commissioned Atacama Pathfinder EXperiment (APEX)\footnote{This
  publication is based on data acquired with the Atacama Pathfinder
  Experiment (APEX). APEX is a collaboration between the
  Max-Planck-Institut f\"ur Radioastronomie, the European Southern
  Observatory, and the Onsala Space Observatory.} is located,
currently is the only site able to perform routine observations within
the high frequency atmospheric windows at high ($\leq$10$''$) spatial
resolutions. The CHAMP$^+$ instrument, developed by the MPIfR and SRON
Groningen, is the only instrument in the world able to simultaneously
observe molecular line emission in the 650 and 850 GHz atmospheric
windows on sub-arcminute spatial scales and is thus ideally suited to
probe the warm gas directly through observations of the 6--5, 7--6 and
8--7 transitions of CO and its isotopologues with 7--9$''$ angular
resolution \citep{Kasemann06,Guesten08}.  CHAMP$^+$ has 14 pixels (7
in each frequency window) and is thus capable of fast mapping of the
immediate surroundings of embedded YSOs.  The {\it Herschel Space
  Observatory} will allow observations of far-IR CO lines at spectral
and spatial resolution similar to APEX ($\sim$10$''$).  Note also that
the beam of Herschel is comparable or smaller than the field of view
of CHAMP$^+$ at its longer wavelenghts ($\sim$500$\mu$m). Thus the
CHAMP$^+$ data obtained here provide information of the distribution
of warm gas within the Herschel beams.

In paper I, we presented the results for one source, HH 46 IRS. In this
paper, we present observations of CO and its isotopologues using
CHAMP+ for seven additional embedded YSOs and compare them with the HH
46 case. $\S$2 presents the sample and observations. The resulting
spectral maps are shown in $\S$3. $\S$4 discusses the envelope and
outflow structure, while the heating within protostellar envelopes and
molecular outflows is analyzed in $\S$5. $\S$6 investigates the
relation between the emission of more complex molecules and the
emission of high-$J$ CO. Final conclusions are given in $\S$7.

\placeTablePaperSevenOne

\placeTablePaperSevenTwo

\placeTablePaperSevenThree
\placeTablePaperSevenFour

\section{Sample and observations}

\subsection{Observations}
The sample was observed using the CHAMP$^+$ array \citep{Kasemann06,Guesten08} on APEX
\citep{Kasemann06}. CHAMP$^+$ observes simultaneously in the 650 GHz
(450 $\mu$m, CHAMP$^+$-I) and 850 GHz (350 $\mu$m, CHAMP$^+$-II)
atmospheric windows.  The array has 7 pixels for each frequency,
arranged in a hexagon of 6 pixels around 1 central pixel, for a total
of 14 pixels. During the observations, the backend consisted of 2 Fast
Fourier Transform Spectrometer (FFTS) units serving the central pixel,
and 12 MPI-Auto-Correlator Spectrometer (MACS) units serving the other
pixels. The FFTS units are capable of observing up to a resolution of
0.04 km s$^{-1}$ (0.12 MHz) and the MACS units to a spectral
resolution of 0.37 km s$^{-1}$ (1 MHz), both at a frequency of 806
GHz.  The observations were done during three observing runs in June
2007, October-November 2007 and July 2008 in three different line
settings; see Table \ref{7:tab:settings}.  Note that in July 2008, all
pixels were attached to FFTS units and no MACS backends were used.
For mapping purposes, the array was moved in a small hexagonal pattern
to provide a fully Nyquist sampled map or in an on-the-fly (OTF)
mode. The $`$hexa$'$ pattern covers a region of about
30$''\times30''$. For some sources, a slightly larger area was mapped
in a small OTF (40$''\times40''$) map in $^{12}$CO 6--5/7--6. Both kinds of maps were regridded to a regular grid with standard rebinning algorithms included in the CLASS package\footnote{CLASS is part of the GILDAS reduction package. See http://www.iram.fr/IRAMFR/GILDAS for more information.}. A binning method that uses equal weight binning was adopted to produce contour maps. A pixelsize of 1/3rd of the beamsize was adopted to create accurate contour maps.  CLASS was also used as the main reduction package for individual lines. At the
edge of the maps, noise levels are often higher due to the shape of
the CHAMP$^+$ array. NGC 1333 IRAS 2, TMR 1 and RCrA IRS 7 were
observed using the hexa mode and have smaller covered areas. Due to
the different beams at 690 and 800 GHz, small differences exist
between the areas covered in CO 6--5 and 7--6. For the C$^{18}$O 6-5
/$^{13}$CO 8--7 (setting C; only done for HH 46), a stare mode was
used to increase the S/N within the central pixel.  A position switch
of 900 $''$ was used for all settings, except for the stare setting C,
which used a beam-switching of 90$''$.

Beam efficiencies, derived using observations on planets during each run, are 0.56 for
CHAMP$^+$-I and 0.43 for CHAMP$^+$-II. These efficiencies were found to vary by less than 10 $\%$ over a single observing run and between runs.  Between pixels the variation is similar, all within 10$\%$. Depending on observing
mode, a given sky position is covered by many (all) pixels, so the average
(relative) calibration error will be a few per cent. Typical single side-band system
temperatures are 700 K for CHAMP$^+$-I and 2100 K for CHAMP$^+$-II.
The 12-m APEX dish produces a beam of 9$''$ at 650 GHz and 7$''$ at
850 GHz. Pointing was checked on various planets and sources and was
found to be within 3$''$. Calibration of the sources was done using
similar observations, as well as hot and cold loads.
The typical sideband rejection at both frequencies was measured to be less than 10dB. At APEX 10 dB is used as the input to the calibrator (both continuum and line. From all these effects, the total calibration error is estimated to be $\sim$30$\%$, including the atmospheric model. 
Note that all technical aspects of CHAMP$^+$ will be discussed more extensively in G\"usten et al. (to be submitted to A\&A, see also \citet{Guesten08}).

  The difference in CO 7--6
emission of 60$\%$ for IRAS 12496-7650 in \citet{vanKempen06} and this
paper can be accounted for by the respective calibration and pointing
errors of FLASH and CHAMP$^+$.  A factor of 2 difference is found with
the observations of TMR~1 in \citet{Hogerheijde98}. Since a similar
factor of 2 is seen between the observations of T Tau in
\citet{Schuster95} and \citet{Hogerheijde98}, it is most likely that
the side-band gain ratios for the observations taken by
\citet{Hogerheijde98} were incorrectly calibrated.

\subsection{Sample}
The sample consists of eight well-known and well-studied embedded
protostars, with a slight bias toward the southern sky.  A variety of
protostars in mass, luminosity, evolutionary stage and parental cloud
is included. All sources have been studied in previous surveys of
embedded YSOs in the sub-mm
\citep[e.g.,][]{Jorgensen02,Jorgensen04a,Groppi07}. More information
on southern sources can be found in \citet{vanKempen06} (IRAS
12496-7650), van Kempen et al. 2009 submitted (IRAS 12496-7650, RCrA
IRS 7 and Ced 110 IRS 4) and Paper I (HH~46).  Table
\ref{7:tab:source} gives the parameters of each source and its
properties. References to previous continuum studies (column 7)
include infrared (IR) and (sub)millimeter dust continuum photometry,
and studies of CO emission (column 8) include both submillimeter and
far-IR lines. Note that the binary source of N1333 IRAS 2A, N1333 IRAS 2B is not covered by our map.

\subsection{Spectral energy distribution}
For all sources, SED information was acquired from the literature,
ranging from near-IR to the (sub-)mm wavelengths. Spitzer-IRAC (3.6,
4.5, 5.6, 8.0 $\mu$m), Spitzer-MIPS (24, 70 and 160 $\mu$m),
submillimeter photometry from JCMT/SCUBA \citep{diFrancesco08},
SEST/SIMBA and IRAM-30m/MAMBO are all included, if available. For
sources with no reliable Spitzer fluxes, IRAS fluxes at 12, 25, 60 and
100 $\mu$m were used. In addition to the obtained fluxes, the
bolometric luminosity and temperature, $L_{\rm{bol}}$ and
$T_{\rm{bol}}$ were calculated using the mid-point method
\citep{Dunham08}. The results can be found in Table
\ref{7:tab:source}. Note that due to a lack of mid- and far-IR fluxes
with sufficient spatial resolution, the derived values for IRAS
12496-7650 and Ced 110 IRS 4 are highly uncertain, with expected
errors of 50$\%$, while no $T_{\rm{bol}}$ and $L_{\rm{bol}}$ could be
determined for RCrA IRS 7 due to confusion. For other sources, errors
are on the order of 10-20$\%$.

\section{Results}
\subsection{Maps}
Figures \ref{7:fig:N1333} to \ref{7:fig:rcra} show the integrated
intensity and the red- and blue-shifted outflow emission mapped in the
CO 6--5 line, as well as the spectra at the central position of all
observed emission lines. Both the integrated intensity and outflow
maps were spatially rebinned to a resolution of 10$''$.  Figures
\ref{7:fig:spec1} to \ref{7:fig:spec4} show the CO 6--5 and 7--6
spectra taken in the inner 40$''\times40''$ of all the sources, also
binned to square 10$''\times10''$ bins.

Table \ref{7:tab:COprop} presents the intensities of both CO
lines. Here, the total integrated emission, the peak main beam
temperature, the emission in the red and blue outflow wings shown in
the outflow maps and the noise levels are given for the central
position, as well as at positions with clear detections of outflow wings away from the center. The limits for the
red and blue outflow were chosen to be -10 to -1.5 km s$^{-1}$ with
respect to the source velocity for the blueshifted emission and +1.5
to +10 km s$^{-1}$ with respect to the source velocity for the
redshifted emission for most sources. These limits were chosen after
examination of the profiles of the spectra and subtracting gaussians
that were fitted to the central 3 km s$^{-1}$ of the profile. For all
sources, except RCrA IRS 7, the difference in the blue- and
red-shifted emission between the two methods was less than 5 $\%$. The
outflowing gas of RCrA IRS 7 was derived by limiting the red- and
blue-shifted emission to -20 to -8 and +8 to +20 km s$^{-1}$ from line
center.  This corresponds to a FWHM$\approx$8 km s$^{-1}$ for the
quiescent material. Broad line widths of 3 km s$^{-1}$ are also seen
for the rarer isotopologues C$^{18}$O and C$^{17}$O, much wider than
typically observed for these species in low-mass YSOs
\citep{Schoeier06}.  Noise levels (see Table \ref{7:tab:COprop})
differ greatly between sources and even within single maps.

\placeFigurePaperSevenOne
\placeFigurePaperSevenTwo

\placeFigurePaperSevenThree 
\placeFigurePaperSevenFour 

$^{12}$CO 6--5 and 7--6 was detected at the central position of all
sources, ranging from 20.8 K km s$^{-1}$ ($T_{\rm{peak}}$=7.8 K) for
Ced 110 IRS 4 for CO 6--5 to 407.6 K km s$^{-1}$ ($T_{\rm{peak}}$=46.3
K) for RCrA IRS 7A for the CO 7--6 line.  All maps show extended
emission, except for IRAS 12496-7650, which shows unresolved emission
in the CO 7--6 transition. However, the scales on which extended
emission is seen varies significantly, with detections at all mapped
positions for RCrA IRS 7 to only 1 or 2 for IRAS 12496-7650, TMR 1 and
Ced 110 IRS 4.  All sources except RCrA IRS 7 and NGC 1333 IRAS 2 show
spectra with a single peak over the entire map, while the latter two
have spectra that are self-absorbed.

Figures \ref{7:fig:spec1} to \ref{7:fig:spec4} clearly identify the
variation of the line profiles across the maps, especially when
outflowing gas is present, such as in the maps of NGC 1333 IRAS 2, BHR
71, HH~46 and RCrA IRS 7. The sources for which little to no shocked
emission is seen do show spatially resolved CO 6--5 and 7--6 emission,
but always quiescent narrow emission located close to the central
pixel.

\subsection{Outflow emission}

From the outflow maps in Fig. \ref{7:fig:N1333} to \ref{7:fig:rcra},
it can be concluded that the contributions from shocks within the
bipolar outflows to the warm gas differ greatly from source to
source. RCrA IRS 7, NGC 1333 IRAS 2, HH~46, and BHR 71 produce
spatially resolved flows, but TMR 1, L~1551 IRS 5, IRAS 12496-7650 and
Ced 110 IRS 4 do not have warm shocked gas that results in broad high-$J$ CO
in any of the off-positions. Small shocks on the source position are
seen, but are generally weak.  For all sources in the sample, outflow
emission has been detected in low ($J_{\rm{up}}\leq3$) excitation CO
lines, although such flows have large differences in spatial scales,
ranging from tens of arcminutes to the central twenty arcseconds
\citep[e.g.][van Kempen et al. 2009
  submitted]{MoriartySchieven88,Fridlund89,Cabrit92,Bachiller94,Bourke97,Hogerheijde98,Parise06}. Table
\ref{7:tab:COprop} gives the integrated intensities of the CO emission
at selected off-positions for the outflowing gas.

\subsection{Isotopologue observations at the central position}

For HH~46, transitions of $^{13}$CO $J$=6--5 and 8--7 and C$^{18}$O
$J$=6--5 as well as [C I] 2--1 were observed (see Table
\ref{7:tab:settings} and \ref{7:tab:source}). TMR~1, Ced 110 IRS 4,
IRAS 12496-7650 and BHR 71 were observed in $^{13}$CO 6--5 and [C I]
2--1. The results at the central position can be found in Fig.
\ref{7:fig:N1333} to \ref{7:fig:rcra}, as well as Table
\ref{7:tab:isoprop}. All $^{13}$CO 6--5 spectra can be fitted with
single gaussians. However, the width of the gaussians varies with a
FWHM of $\sim$1.2 km s$^{-1}$ for HH~46 and BHR 71 to $\sim$2.2 km
s$^{-1}$ for Ced 110 IRS 4 and IRAS 12496-7650.  [C I] 2--1 is
detected for TMR 1, HH 46 and Ced 110 IRS 4 ($\sim$3 $\sigma$). No
line was found for IRAS 12496-7650 and BHR 71 down to a 1$\sigma$
limit of 0.6 K in 0.7 km s$^{-1}$ bins.  Integrated line strengths are
on the order of 3 K km s$^{-1}$ with widths of 0.75 km s$^{-1}$.
$^{13}$CO 8--7 and C$^{18}$O 6--5 were observed towards HH 46 only,
but not detected down to a 1$\sigma$ level of 0.15 and 0.3 in a 0.7 km
s$^{-1}$ channel.

\subsection{$^{13}$CO 6-5 and [C I] 2-1 maps}

For $^{13}$CO 6--5 and [C I] 2--1 maps were also obtained. Spectra at
the central position are given in Figs. \ref{7:fig:N1333} to
\ref{7:fig:rcra}. Integrated intensity maps are presented in
Fig. \ref{7:fig:app_1} ($^{13}$CO 6--5) and Fig. \ref{7:fig:app_2} ([C
  I] 2--1) It is seen that such observations are often dominated by
centrally located unresolved emission, but not always peaked at the
source. Both HH~46 and BHR~71 also show some isotopic emission
associated with the outflow. For HH~46, $^{13}$CO 6--5 is only
detected off-source for the blue outflow, where part of the outflow is
unobscured by cloud or envelope (Paper I).

\placeFigurePaperSevenFive
\placeFigurePaperSevenSix

\placeFigurePaperSevenSeven
\placeFigurePaperSevenEight

\placeFigurePaperSevenNine
\placeFigurePaperSevenTen
\placeFigurePaperSevenEleven
\placeFigurePaperSevenTwelve

\placeFigurePaperSevenThirteen
\placeFigurePaperSevenFourteen

\section{Envelope}

\subsection{Envelope models}
In order to investigate whether high-$J$ CO emission can be reproduced
by a passively heated envelope, the properties of the protostellar
envelopes were calculated by modelling 850 $\mu$m continuum images,
SED information and the 1-D dust radiative transfer code DUSTY
described in \citet{Ivezic97}
\citep[][]{Schoeier02,Jorgensen02}. Similar to previous studies, the so-called OH5 dust opacities were used \citep{Ossenkopf94}.  Although NGC 1333 IRAS 2, L1551
IRS 5 and TMR 1 were included in a similar study by
\citet{Jorgensen02}, new fluxes from Spitzer \citep[e.g.,][ see Table
  \ref{7:tab:source}]{Luhman08,Gutermuth08} have since come available
and must be included.  To constrain the models both the SED from
$\sim$50 $\mu$m to 1.3 mm and the spatial distribution of the sub-mm
continuum emission at 850 $\mu$m were used, with the exception of BHR
71 and Ced 110 IRS 4, for which no 850 $\mu$m map is available. A
normalized radial emission profile with a power law index of 1.5 for the 850 $\mu$m emission is
assumed for these two sources.  The radial profiles of L 1551 IRS 5,
RCrA IRS 7A, NGC 1333 IRAS 2 and TMR 1 were obtained from 850 $\mu$m
images of the processed SCUBA archive \citep{diFrancesco08}. For HH~46
and IRAS 12496-7650, recent data from LABOCA at 870 $\mu$m were used
(see Paper I and Nefs et al. in prep.).

The parameters of the best-fitting envelope models can be found in
Table \ref{7:tab:env}, together with the corresponding physical
parameters of the envelope. The three main parameters of DUSTY, $Y$,
the ratio over the inner to outer radius, $p$, the power law exponent
of the density profile $n \propto r^{-p}$\footnote{Note that this power law index is not the same as the power law index $p$ for the normalized 850 $\mu$m radial emission profile} and $\tau_{100\mu\rm{m}}$,
the opacity at a 100 $\mu$m, are scaled by the $L_{\rm{bol}}$ and the
distance, $D$ given in Table \ref{7:tab:source}, to get the physical
properties of each source.  Fig. \ref{7:fig:dusty} shows the
best-fitting model of each source of the radial profiles of the 850
$\mu$m images and the SEDs of the entire sample.

\placeTablePaperSevenFive

The inner radii of the protostellar envelopes range from 5 to 35 AU corresponding to $T_{\rm{d}}=250$ K, a limit chosen by us.  
Most sources also show a steep profile with all sources having $p\geq1.7$, with the exception of Ced 110 IRS 4, which has a $p$=1.4.

\subsection{CO emission within protostellar envelopes}

Using the best-fit envelope temperature and density structure derived
from the dust emission, the CO intensities and line profiles from the
protostellar envelopes were in turn simulated with the self-consistent
1D molecular line radiative transfer code RATRAN \citep{Hogerheijde00}
using data files of the LAMDA database \citep{Schoeier05}. CO
abundances with respect to H$_2$ are taken from
\citet{Jorgensen05a}. A `drop' abundance with
$X_0$=2.7$\times$10$^{-4}$ and $X_d$=10$^{-5}$ is adopted. This `drop'
abundance profile describes a warm ($T>T_{\rm{freeze}}$) inner region
with a high abundance $X_0$ and a region in which $T<T_{\rm{freeze}}$
and $n>n_{\rm{de}}$ where CO is frozen out to a low abundance
$X_d$. In the outer region ($n<n_{\rm{de}}$), the abundance is again
high at $X_0$ because freeze-out timescales become longer than the
typical life-times of protostars.  In our models, $T_{\rm{freeze}}$=30
K and $n_{\rm{de}}$=10$^5$ cm$^{-3}$ are adopted, following the
conclusions of \citet{Jorgensen05a}. There, the derived abundances are
based upon the emission of low-excitation optically thin lines such as
C$^{18}$O lines (both 2--1 and 3--2, 1--0 is often dominated by the
very cold cloud material) and C$^{17}$O. Paper I showed
that contrary to the low-$J$ CO lines there is little difference
between 'jump' and 'drop' abundances for the emission in high-$J$
transitions. A static velocity field is assumed with a turbulent width
of 1 km s$^{-1}$.  Due to the static nature of the velocity field,
excessive self-absorption is seen in line profiles of CO lines up to
the 8--7 transition. The total area of a gaussian fitted to the line
wings is used in those cases to derive an upper limit. This is a very
strict limit as the true CO emission associated with the envelope is
best fitted by a infall velocity \citep{Schoeier02}, producing
integrated intensities between the two limits. However,
\citet{Schoeier02} show that the envelope emission modelled with an
infall velocity is in the worst case only a factor 2 greater than the
intensity derived from the static envelope. The lower limit is derived
from the actual modelled integrated intensity with the static velocity
field.  

\placeFigurePaperSevenFifteen

\placeFigurePaperSevenSixteen

 Figure \ref{7:fig:CO_model} shows the resulting integrated
 intensities produced by the model protostellar envelopes of all CO
 lines from $J$=1--0 up to $J$=19--18. The data for all transitions
 were convolved with a 10$''$ beam used for CHAMP$^+$, except the
 three lowest transitions, which are convolved with a beam of
 20$''$. Such beams are typical for single-dish submillimeter
 telescopes. 10$''$ will also be the approximate beam for several
 transitions covered by the {\it PACS} and {\it HIFI} (Band 6 and 7)
 instruments on Herschel at the higher frequencies.  Overplotted are
 the observed line strengths from various CO lines of different
 studies, including the far-IR high-$J$ CO lines of \citet{Giannini99}
 and \citet{Giannini01}, assuming that all flux observed by the
 ISO-LWS in its 80$''$ beam originates in a 10$''$ region. See Table
 \ref{7:tab:source} for the references used.

\placeTablePaperSevenSix
\placeTablePaperSevenSeven

The low-$J$ CO emission can often be completely reproduced by the
envelope models, as it is dominated by the colder gas in the outer
regions of the envelopes.  For HH~46, Ced 110 IRS 4 and RCrA IRS 7,
the observed quiescent emission in the CO 6--5 and 7--6 lines is
clearly brighter by a factor of 3--5 than the modelled envelope
emission, even for the $^{13}$CO 6--5 lines (see Table
\ref{7:tab:13CO}).

For RCrA IRS 7, the difference is almost an order magnitude.  Deep
C$^{18}$O 6--5 spectra are needed to fully pin down the envelope
models.  
In Paper I, the quiescent CO 6--5 and 7--6 emission of
HH~46 was attributed to 'photon heating' \citep{Spaans95}, both by UV
from the accretion disk and shocks inside the outflow
cavity. Fig. \ref{7:fig:CO_model} clearly shows that this method of
heating likely applies to other sources.

In contrast, the envelopes of both IRAS 12496-7650 and L~1551 IRS 5
can account for all the emission detected in the $^{12}$CO 6--5 and
7--6 lines.  \citet{Giannini99} and \citet{Giannini01} report emission
of high-$J$ CO (14--13 to 19--18) lines at far-IR of NGC 1333 IRAS 2,
IRAS 12496-7650 and RCrA. For RCrA, RCrA IRS 7 is within the beam, but
the emission is probably dominated by emission from RCrA itself.
Fluxes in excess of $10^{-20}$ W cm$^{-2}$ are seen for IRAS
12496-7650. They assumed that the flux originates within the central
3$''$ ($\sim$400 AU). It is very clear from Fig. \ref{7:fig:CO_model}
that such observed emission cannot be produced by a passively heated
envelope. A more likely explanation is that the CO emission detected
with ISO-LWS is either located outside the inner 10$''$ as is the
likely case for IRAS 12496-7650 (see \citet{vanKempen06}) or
associated with an energetic outflow.

\section{Outflows}
\subsection{Shocks}

\citet{Bachiller99} propose an evolutionary sequence of outflows
around low-mass protostars, with young deeply embedded YSOs (Class 0)
producing highly collimated and energetic outflows, and with more
evolved embedded YSOs (Class I) producing outflows which show less
energetic shocks and a wider opening angle \citep[see
  also]{arce07}. Observations of the shocked $^{12}$CO 6--5 and 7--6
gas in the outflow directions confirm the scenario that the shocks
within the vicinity of the protostar grow weaker in energy over
time. The Class 0 sources (NGC 1333 IRAS 2, BHR 71 and RCrA IRS 7) all
show shocked warm gas in their CO 6--5 and 7--6 lines, in both blue-
and red-shifted outflow. Of the Class I sources, only HH 46 shows
shocked gas in its red-shifted outflow.  L~1551 IRS 5, one of the
most-studied molecular outflows
\citep{MoriartySchieven88,Bachiller94}, shows little emission in the
high-$J$ lines associated with outflow shocks. Although several
spectra do show a small outflow wing, the integrated emission in the
wings is not higher than a few $\sigma$. All other Class I flows show
no sign of shocked warm gas. HH~46 and L~1551 IRS 5, even though
classified as Class I, have massive envelopes of a few M$_\odot$, more
characteristic of Class 0 \citep{Jorgensen02}. L~1551 IRS 5 is
believed to be older and to consist of several sucessive ejection events
\citep{Bachiller94,White00}.  

Figure \ref{7:fig:vmax} shows the maximum outflow velocities of the $^{12}$CO 6--5 and 3--2 lines vs. the bolometric temperature, envelope mass and outflow force. A clear absence of warm high-velocity material is seen for sources with a higher bolometric temperature. Only the cold outflow  of IRAS 12496-7650 is seen at higher $T_{\rm{bol}}$ \citep{vanKempen06}. Similarly, there is also a clear relation between the mean outflow force of both red and blue outflow lobes, and the maximum velocity seen in both $^{12}$CO 3--2 and $^{12}$CO 6--5 emission.  Outflow forces are derived from the spatial scales and velocities of low-excitation CO  line emission (1--0, 2--1 and 3--2) \citep[Paper I, van Kempen et al. 2009 submitted]{Cabrit92,Bourke97,Hogerheijde98}.  

\placeFigurePaperSevenSeventeen

\placeFigurePaperSevenEighteen

\subsection{Temperatures of the swept-up gas}
The excitation temperature of the outflowing swept-up gas can be
derived from the ratios of different CO line wings. As an example,
Fig. \ref{7:fig:overplot} shows the CO 3--2 data from van Kempen et
al. 2009 submitted and Paper I and CO 6--5 data from this
paper overplotted on the same scales for a few sources. The CO 6--5
data have not been binned to the larger CO 3--2 beam, so the
comparison assumes similar volume filling factors of the shocked gas;
for HH 46 in paper I it has been checked that rebinning gives similar results
within the uncertainties. If the density is known this excitation
temperature can be related to the kinetic temperature using the
diagnostic plots produced by the RADEX radiative transfer code
\citep{vanderTak07}. 


As can be seen from the temperature analysis of HH~46
(Paper I), there are many uncertainties, leading to error bars as large as 50 K on the inferred temperatures. The ratios depend
on the velocity, the optical depth of the line wings and the ambient
density at different distances from the source. For HH~46, a drop in
temperature was observed if the density remains constant, but 
it is more plausible that the density is lower at larger radii which
will result in a constant kinetic temperature with distance.

Table \ref{7:tab:temp} gives the median temperatures using the extreme
velocities of the line wings with intensities $>3\sigma$, assuming the
line wings are optically thin and with an ambient cloud density of
10$^{4}$ cm$^{-3}$. 
CO 3--2 spectra can be found in
\citet{Hogerheijde98} (TMR~1, L~1551 IRS 5), \citet{Knee00} (NGC 1333
IRAS 2), \citet{Parise06} (BHR~71), \citet{vanKempen06} (IRAS
12496-7650), van Kempen et al. 2009 submitted (RCrA IRS 7, Ced 110 IRS
4) and Paper I (HH~46).

The main error on these temperature estimates is the optical depth in
the outflow wings. Deep $^{13}$CO 3--2 observations show that outflows
can be optically thick. \citet{Hogerheijde98} find optical depths in
the $^{12}$CO 3--2 line wings, $\tau_{\rm{wing}}$, on the order of 10,
while in Paper I $\tau_{\rm{wing}}$=1-1.8 is found for
HH~46. Even if the optical depths are assumed to be similar in CO 3--2
and CO 6--5, kinetic temperatures can be almost 50$\%$ lower than
those given in Table \ref{7:tab:temp}. 

For a more thorough discussion about the density of the surrounding envelope and cloud material, see Paper I. Our choice of the density of 10$^4$ cm$^{-3}$ is based on the typical densities found in model envelopes of  \citet{Jorgensen02} at distances of a few thousand AU, corresponding to the 10 K radius where the envelope merges with the surrounding molecular cloud. Higher densities will produce lower outflow temperatures (see paper I for more extensive discussion.

 A theoretical study by
\citet{Hatchell99} using a jet-driven bow shock model suggests that
the kinetic temperatures of molecular outflows are typically of order
of 50-150 K along the axis and rising toward the location of the bow
shock, with overall values increasing with higher jet velocities.  The
temperatures found in Table \ref{7:tab:temp} agree well with
these predicted temperatures. For several outflows, e.g., IRAS
12496-7650 (blue) or Ced 110 IRS 4 (red), temperatures are
significantly lower, however.  The lowest outflow temperatures are found for
the flows of L~1551 IRS 5, where
CO 4--3/6--5 ratios are on the order of 8
or higher. At an assumed density of 10$^4$ cm$^{-3}$, this corresponds
to kinetic temperatures of 50 K and lower. If both line wings are
optically thin, densities must be lower than 10$^{3}$ cm$^{-3}$ to
produce temperatures of $\sim$100 K, observed for other flows.

\section{Heating processes in the molecular outflow and protostellar envelope}

As discussed in $\S$ 4, several sources show quiescent, narrow
$^{12}$CO 6--5 and 7--6 emission that is more intense than can be
produced by an envelope model. Moreover, strong narrow high-$J$ CO
emission is observed off-source along outflow axes for most sources.
In Paper I, we proposed that for HH~46 the quiescent
narrow $^{12}$CO 6--5 and 7--6 line emission originates within the
outflow cavity walls heated to 250--400 K by an enhancement factor
$G_0$ with respect to the standard interstellar radiation field of a
few hundred. The UV photons can be created by jet shocks in the
outflow cavities as well as in the disk-star accretion boundary layer
near the central protostar. This heating method was first proposed by
\citet{Spaans95}, but extended by Paper I to both the
inner envelope as well as the outflow cavity walls much further from
the central star. The data presented in this paper show that photon
heating is present in other protostars as well, especially in outflow
cavities. Even in sources with little to no outflow, such as Ced 110
IRS 4, relatively strong narrow $^{12}$CO 6--5 and 7--6 lines are seen
at positions not associated with the central protostar, see
Fig. \ref{7:fig:CO_model}. The presence of significant radiative
`feedback' from the protostar on its surroundings may have
consequences for the collapse of the envelope, limiting the accretion
rate and mass of the star and suppressing disk fragmentation
\citep[e.g.,][]{Offner09}.

The origin of such quiescent high-$J$ CO emission at the source
velocity is physically different from both the thermal emission of the
protostellar envelope (see $\S$ 6.4), emission from shocks present at
the working surfaces of outflows \citep{Reipurth99,Raga07}. 

Slow ($v_s=5-10$ km s$^{-1}$) C-shocks may produce similar quiescent
levels of CO 6--5 and 7--6 emission in outflows
\citep{Draine84,Spaans95}. However, the presence of quiescent
$^{12}$CO 6--5 and 7--6 emission in the envelopes of TMR~1 and Ced 110
IRS 4, both of which show little to no spatially resolved outflow
emission in the CO 3--2 line \citep[ and van Kempen et
  al. submitted]{Hogerheijde98}, is more easily explained with the
photon heating scenario than with slow C-shocks.  In addition, the
narrow line widths for other sources argue against this scenario.

\subsection{Envelope and outflow of BHR 71}

The proposed model that photon heating takes place both in the outflow
cavities and the inner envelope is clearly illustrated by the
observations of BHR 71. At the north and south position of the
outflow, shocked gas is clearly detected, but almost no shocked gas is
seen near the central star. Quiescent gas with $\Delta V$=1.5 km
s$^{-1}$ is observed both within the inner envelope and along the
outflow.  In the $^{12}$CO 7--6 map, the photon heating in the
envelope can be strongly identified by the central contour.  BHR~71 is
the only source for which the quiescent emission at the outflow
position is (much) brighter than that seen at the position of the
envelope. It is likely that strong shocks producing copious UV are
present in the main outflow.  This outflow \citep{Bourke97} is
clearly detected at the edges of the map, but the secondary weaker
outflow, associated with the IRS2 position, as seen by
\citet{Parise06} is not detected down to the noise levels. Shocked CO
6--5 and 7--6 emission is also only seen at relatively low levels in
the inner 20$''$ and strongly increases at 40$''$ away from the source
for both outflows.

\subsection{The 'fossil' outflow of L~1551 IRS 5}

The outflow of L~1551 IRS 5 has been considered an example of an older
outflow, due to its large size \citep{MoriartySchieven88,Fridlund89},
large opening angle \citep{Bachiller94} and other submillimeter
properties \citep{Cabrit92,Bachiller94,Hogerheijde98}. \citet{White00}
constructed a detailed model from a wide variety of space- and
ground-based IR continuum and spectroscopic observations, including a
wide opening angle of the outflow ($\sim$50$^{\circ}$) and low
densities inside the cavity. This view is clearly confirmed by the low
derived temperatures of 40 to 50 K of the shocked gas, which is much
cooler than in other sources. Little high-$J$ CO emission is seen at
the off-positions, and the maps are dominated by the emission of the
central source. Comparison of the peak temperatures of CO 4--3
\citep{Hogerheijde98}, 6--5 and 7--6 shows that $T_{\rm{peak}}$ at
high-$J$ transitions is similar to that at low-$J$ CO transitions. This suggest that even though the outflow is present, the density in the surrounding cloud must be quite low.

\subsection{The PDR of RCrA IRS 7}

The lines of RCrA IRS 7 are an order of magnitude stronger than those
of equivalent sources in different clouds. Even at off-positions at
the edge of the observed area, integrated intensities larger than 300
K km s$^{-1}$ are seen. Even with the high luminosity of $\sim$20-30
L$_\odot$, these integrated intensities cannot originate from heating
by RCrA IRS 7 itself. A likely explanation is the proximity of the A5
star RCrA ($L_{\rm{bol}}=130$ $L_\odot$) at 36 $''$ (4,500 AU). RCrA
was observed by ISO-LWS, and many strong high-$J$ CO lines (CO 14--13
to 21--20) were detected in the large beam of ISO
\citep{Lorenzetti99,Giannini99}. Models were put forward with the
emission originating in relatively small ($\sim$0.002 pc), dense
($>$10$^{6}$ cm$^{-3}$) and hot ($T>300-1000$ K) regions
\citep{Giannini99}.  However, the spatial distribution of the CO 6--5
and 7--6 seen in Fig. \ref{7:fig:rcra} does not agree with this
hypothesis. A much more likely explanation is that RCrA itself
irradiates the outer edges of the cloud and envelope around RCrA IRS
7a, creating a PDR at its surface, much like the case of intermediate
mass sources in Orion \citep{Jorgensen06}.

\subsection{Presence of [C I] 2--1}

A limit on the amount of FUV/X-ray emission that is available to
dissociate CO and H$_2$ can be derived from the presence of the [C I]
2--1. Photodissociation of CO can only
occur at 912-1100 \AA \ \citep{vanDishoeck88}, so the absence of
strong [C I] 2--1 emission in the outflow cavities of most protostars
suggests that the radiation field in outflows does not produce
sufficiently energetic radiation, but still heats the cavity walls to
a few hundred K.  This would also limit the shock velocities to $<$90
km s$^{-1}$, since higher shock velocities produce CO and H$_2$
dissocating photons \citep{Neufeld89}.

[C I] is detected in the inner 10$''$ for HH~46, TMR 1 and Ced 110 IRS
4 at levels of 2 K km s$^{-1}$. For IRAS 12496-7650 and BHR~71, no [C
  I] 2--1 was detected. The low emission in all sources can be
accounted for by a C abundance of a few times 10$^{-6}$ with respect
to H$_2$. Such abundances can be maintainted by low UV levels produced
by cosmic ray radiation \citep{Flower94}.

\section{Conclusions}

In this paper, we have presented the first $^{12}$CO 6--5 and 7--6
maps of a sample of 8 low-mass protostars with a large range of
luminosities, evolution and densities, as well as several isotopologue
and [C I] observations. All observations have been carried out with
the CHAMP$^+$ instrument.  The main conclusions of this paper are
\begin{itemize}

\item Warm gas, as traced by $^{12}$CO 6--5 and 7--6, is present in
  all observed protostars at the central position.  Three different
  origins of the warm gas emission are found: (i) the inner envelope
  heated passively by the protostellar luminosity; (ii) shocked gas in
  the outflow; and (iii) quiescent gas heated by UV photons. This
  latter mechanism, as detailed in \citet{Spaans95} and in Paper I,
  generally dominates the extended high-$J$ CO emission.

\item Envelope models show that for HH~46 and Ced 110 IRS 4, passive
  heating of the envelope is insufficient to explain the observed
  $^{12}$CO 6--5, 7--6 and $^{13}$CO 6--5 lines, requiring heating of
  the envelope by UV photons even at the (0,0) position.

\item Photon heating of the cavity walls takes place on arcmin scales
  for the outflow cavities of several other Class 0 and Class I
  protostars, as seen at positions off-source of BHR 71, NGC 1333 IRS
  2, HH~46 and L~1551 IRS 5. The necessary UV photons are created by
  internal jet shocks and the bow shock where the jet interacts with
  the ambient medium, in addition to the disk-star accretion boundary
  layer. The distribution of the quiescent CO 6--5 and 7--6 emission
  of BHR~71, with narrow emission stronger at larger distances from
  the source, confirms the hypothesis that UV photons necessary for
  the heating can originate both mechanisms. This suggests that photon
  heating is present in all outflows.

\item The lack of [C I] 2--1 emission in the outflows constrains the
  production of energetic CO dissociative photons in the shocks. The
  observed [C I] emission at the source position can be accounted for
  by a low atomic C abundance that is maintained by cosmic ray induced
  UV photodissocation of CO.

\item Shocked $^{12}$CO 6--5 and 7--6 gas only exists in large
  quantities in flows of more massive sources (NGC 1333 IRAS 2, BHR
  71, HH~46 and RCrA IRS 7), while low-mass flows do not show much
  shocked gas that emits in the $^{12}$CO 6--5 and 7--6 lines. Within
  the proposed model of \citet{Bachiller99}, the outflows of Class I
  are more evolved and are driven with much less energy, producing
  much weaker shocks.  This is also reflected in the decreasing
  maximum velocity of $^{12}$CO 6--5 with lower outflow force.

\item Kinetic temperatures of $\sim$100 K are found for the molecular
  gas in most flows studied here. Such temperatures agree with
  expected conditions of jet driven flows modelled by
  \citep{Hatchell99}. Only the flows of L~1551 IRS 5 and IRAS
  12496-7650 are much colder ($<$50 K), in agreement with the
  `fossil', empty nature of the flows.

\item The very strong intensities at all positions of RCrA IRS 7, an
  order of magnitude higher than can be produced by passive heating,
  must be due to a significant PDR region near the source. Likely,
  RCrA itself heats the outside of the RCrA IRS 7 envelope and cloud
  region.

\end{itemize}

Spatially resolved observations with APEX-CHAMP$^+$ of high-$J$ CO and
isotopologues clearly provide unique insight into the structure and
physical processes of low-mass protostars, as well as their feedback
on the surroundings. The ubiquitous photon heating of the gas found
here may have some consequences for the amount of envelope gas
that can eventually collapse. Detailed multi-dimensional radiative transfer modelling, including outflow cavities  are needed to directly adress this question. Such models are now being developed for high-mass YSOs \citep{Bruderer09}, and will be extended to low-mass YSOs. The UV radiation will also significantly
affect the chemistry along the outflow walls, dissociating molecules
like H$_2$O to OH and O and HCN to CN.
In the future, the Herschel Space Observatory will be able to map even
higher $J$ CO lines as well as H$_2$O, OH and [O I] lines in the
far-IR between 60 and 600 $\mu$m with the {\it PACS} and {\it HIFI}
instruments. Together, radiative transfer modelling, these data sets and future interferometric
observations with ALMA will be able to fully characterize the physical
and chemical structure of the earliest, deeply embedded stages of star
formation.


\begin{acknowledgements}

 TvK and astrochemistry at Leiden Observatory are supported by a
 Spinoza prize and by NWO grant 614.041.004. The staff at APEX and
 many members of the Bonn submillimeter group are thanked for the
 excellent support during observations. CHAMP$^+$ is constructed with
 funds from NWO grant 600.063.310.10. We thank Ronald Stark for
 continued support of the CHAMP$^+$ project. 
\end{acknowledgements}
\bibliographystyle{../../../bibtex/aa}
\bibliography{../../../biblio}

\end{document}